\documentclass[10pt,journal]{IEEEtran}
\usepackage{latexsym}
\usepackage{amsfonts}
\usepackage{amsbsy}
\usepackage{amsmath,amssymb}
\usepackage{times}
\usepackage{graphicx}
\usepackage{enumerate}
\usepackage[usenames]{color}
\usepackage[dvips]{pstcol}
\usepackage{epstopdf}
\usepackage{amsmath}
\usepackage{subfig}
\usepackage{bm}
\usepackage{booktabs}
\usepackage{cite}
\usepackage{color}
\usepackage{xcolor}
\usepackage{algorithm}
\usepackage{algorithmicx}
\usepackage{algpseudocode}
\usepackage{setspace}
\usepackage{bm}
\usepackage{tabularx}
\usepackage{amstext,epsfig,psfrag,graphicx,cite,footnote}
\usepackage{siunitx}
\usepackage{mathrsfs}
\usepackage{booktabs}
\usepackage{multirow}
\usepackage{stfloats}

\ifCLASSINFOpdf
\else
\fi
\begin{document}
\title{Learned Conjugate Gradient Descent Network for Massive MIMO Detection}



%Telephone: (800) 555--1212, Fax: (888) 555--1212}


\author{Yi Wei, Ming-Min Zhao, Mingyi Hong, Min-Jian Zhao and Ming Lei\vspace{-1.5em}
\thanks{Y. Wei, M. M. Zhao, M. J. Zhao and M. Lei are with College of Information Science and Electronic Engineering, Zhejiang University, Hangzhou 310027, China
(email: \{21731133, zmmblack, mjzhao, lm1029\}@zju.edu.cn)

M. Hong is with the Department of Industrial and Manufacturing Systems Engineering, Iowa State University, Ames, IA 50011, USA. (email: mingyi@iastate.edu)

%M. Hong is with the Department of Electrical and Computer Engineering, University of Minnesota, Minneapolis, MN 55455, United States
%(email: mhong@umn.edu)

Part of this paper has been accepted by IEEE ICC 2020.
}}

\maketitle

\begin{abstract}
In this work, we consider the use of model-driven deep learning techniques for massive multiple-input multiple-output (MIMO) detection. Compared with conventional MIMO systems, massive MIMO promises improved spectral efficiency, coverage and range.
Unfortunately, these benefits are coming at the cost of significantly increased computational complexity. To reduce the complexity of signal detection and guarantee the performance, we present a learned conjugate gradient descent network (LcgNet), which is constructed by unfolding the iterative conjugate gradient descent (CG) detector. In the proposed network, instead of calculating the exact values of the scalar step-sizes, we explicitly learn their universal values. Also, we can enhance the proposed network by augmenting the dimensions of these step-sizes. Furthermore, in order to reduce the memory costs, a novel quantized LcgNet is proposed, where a low-resolution nonuniform quantizer is used to quantize the learned parameters. The quantizer is based on a specially designed soft staircase function with learnable parameters to adjust its shape.
 Meanwhile, due to fact that the number of learnable parameters is limited, the proposed networks are easy and fast to train. Numerical results demonstrate that the proposed network can achieve promising performance with much lower complexity.
\end{abstract}


% \fi
\IEEEpeerreviewmaketitle

\section{Introduction}
Massive multiple-input multiple-output (MIMO), which has attracted much attention from both academia and industry, is a promising technology to potentially achieve higher spectral efficiency over existing (small-scale) MIMO systems \cite{6375940}, \cite{6736761}. The main idea of massive MIMO is to equip the transmitter or receiver with a large number of antennas, however this also brings unbearable pressure to signal detection in terms of computational complexity. Therefore, efficient massive MIMO detection algorithms with low complexity and good bit error rate (BER) performance play important roles in the receiver design.
\subsection{Literature Review on Massive MIMO Detection}
Generally, the maximum likelihood (ML) detector is considered to be optimal, but it requires an exhaustive search on all the combinations of transmit symbols, which exhibits exponential computational complexity.
%Therefore, near-optimal algorithms are usually preferred, e.g., the approximate message passing (AMP) detector\cite{7282651}, the semidefinite relaxation (SDR) detector\cite{4475373}, \cite{5447068}, etc.
{Therefore, near-optimal algorithms are usually preferred, e.g., the sphere decoding (SD) detector \cite{1468474} attempts to search over the lattice points that lie in a
certain sphere around the given vector, thereby reducing the search space and the required computation complexity. However, its worst-case complexity is still exponential in $N_t$.} Fortunately, besides higher peak data rates, enhanced link reliability and improved coverage,  theoretical results also indicate that simple, low-complexity and energy-efficient detection algorithms exist when the number of antennas approaches infinity \cite{5595728}, \cite{7244171}.
Some linear detectors, such as the zero forcing (ZF) detector and the linear minimum mean squared error (LMMSE) detector, have been proved to be near-optimal for massive MIMO systems \cite{6375940}. However,  they still require the complex matrix inversion operation. To further reduce the computational complexity, detectors based on truncated Neumann series expansion were proposed in \cite{6572301,7401703,6638142}, where the matrix inversion operation is transformed into a series of matrix-vector multiplications. {In \cite{6954512}, the authors provided a Gauss-Seidel (GS)-based detector to iteratively
approach the performance of the LMMSE detector without the complicated matrix inversion for large-scale MIMO systems.} In \cite{7037382,7024849,7868573}, the conjugate gradient descent (CG) algorithm was employed to iteratively achieve the performance of the LMMSE detector with lower complexity. It was shown in \cite{7037382} that the CG detector outperforms those based on truncated Neumann series expansions in terms of both BER performance and computational complexity. Except for the ML detector, few works were reported in the literature to outperform the LMMSE detector for large scale MIMO systems.
%{\color{blue}These aforementioned signal detection algorithms can be considered ``model-driven'' in the sense that they use some sort of expert knowledge and prior information. }
%In the literature, there is few detectors outperforming MMSE detector for massive MIMO system but ML detector.
\subsection{Background on Deep Learning}
As a popular approach to artificial intelligence, deep learning (DL) has revolutionized many fields, e.g., computer vision and natural language processing, and it has been widely applied to solve wireless physical layer communication problems recently \cite{8214233}.
% DL based methods can be grossly divided into two
%subcategories: 1) data-driven methods, which treat the network as a black box and train it by a huge volume of data, e.g., deep neural networks (DNNs) for channel encoding/decoding \cite{7926071},  channel estimation \cite{8052521}, modulation recognition \cite{8445938} and channel state information (CSI) feedback \cite{8322184}, etc., 2) model-driven methods, which are constructed based on known domain knowledge and physical mechanism, such as %learned denoising-based AMP (LDAMP) \cite{8353153} and
%the LampResNet for mmWave channel estimation \cite{8715473} and the OFDM-autoencoder \cite{8445920}.
%  In particular, in order to construct a network, model-driven DL methods usually unfold {\color{blue}an} existing iterative algorithm, add some adjustable parameters and then train these parameters by DL. For instance, in \cite{Borgerding2016Onsager} and \cite{7442798}, the AMP and iterative thresholding (ISTA) algorithms were improved by unfolding their iterations into networks and learning the corresponding parameters.
{
 Among the various architectures of DL, the deep neural network (DNN) is one of the most effective and promising techniques and it has been used in many applications, e.g., DNNs for channel encoding/decoding \cite{7926071},  channel estimation \cite{8052521}, modulation recognition \cite{8445938} and channel state information (CSI) feedback \cite{8322184}. However, these networks are usually trained as black boxes, and it is difficult to understand their operational mechanisms. Therefore, it is not easy to figure out how to modify their architectures to achieve better results and how to incorporate prior knowledge about the considered problem. In order to address this difficulty, the model-driven DL methods are becoming increasingly popular as a promising alternative  \cite{Hengtao2018}, e.g.,  the LampResNet for mmWave channel estimation \cite{8715473} and the OFDM-autoencoder \cite{8445920}, etc.  This kind of methods blend the internal structure of certain
model-based algorithms with the remarkable power of the state-of-the-art DNN, allowing inference to be performed with a fixed number of layers to achieve optimized performance. In particular, unfolding a well-understood iterative algorithm (also
known as deep unfolding \cite{Hershey2014}) is one of the most popular and powerful techniques to build such a model-driven DL network. For instance, in \cite{Borgerding2016Onsager} and \cite{7442798}, the approximate massage passing (AMP) and iterative shrinkage/thresholding algorithm (ISTA) were improved by unfolding their iterations into networks and learning the corresponding parameters.}

Given the promising advantages of model-driven DL methods, they have also been applied to MIMO detection recently \cite{abs180507631,Jin2018Deep,abs-1812-01571,8646357,Tan2018Improving}. Specifically, the detection network (DetNet) was proposed in \cite{abs180507631} by mimicking a projected gradient descent (PG) like solution for the maximum likelihood optimization. It was shown in \cite{abs180507631} that DetNet achieves a comparable performance to those of the SDR and AMP detectors but with less detecting time. This performance improvement was achieved at the expense of high computational complexity in the offline training process, which took about three days on a standard Intel i7-6700 processor.
%Besides, the work \cite{abs180507631} did not offer the detection performance in massive MIMO system, and when the scale of MIMO is huge, the increased computational complexity would become an inevitable problem in both training and detecting process.
The works \cite{Jin2018Deep} and \cite{abs-1812-01571} also applied this idea to massive overloaded MIMO detection and multi-level MIMO detection, respectively, and comparable detection performance to those of the state-of-the-art algorithms was achieved. In \cite{8646357} and \cite{Tan2018Improving}, DL based MIMO detection networks were proposed by unfolding the orthogonal AMP (OAMP) and belief propagation (BP) algorithms, respectively, and they were also demonstrated to significantly outperform the original algorithms by learning the parameters from a large number of training data.
 %Given the promising advantages of model-driven DL methods, they have also been applied to MIMO detection recently \cite{abs180507631,Jin2018Deep,8646357,Jin2018Deep2}. Specially, the detection network (DetNet) \cite{abs180507631} was proposed by unfolding the iterations of the projected gradient descent (PG) algorithm. The results obtained by the existing algorithm are employed as an initial point of the DetNet and the variables within are optimized based on backpropagation. It was shown in \cite{abs180507631} that DetNet achieves a comparable performance to that of SDR and AMP but with faster speed. For massive overloaded MIMO channels, \cite{Jin2018Deep} presented a trainable PG (TPG)-detector, which improved the DetNet by reducing the trainable parameters significantly but achieved comparable detection performance to those of the
%known algorithms for addressing the same problem. In \cite{8646357}, the Orthogonal AMP (OAMP)-net was proposed by unfolding the OAMP algorithm, and it was demonstrated to significantly outperform the OAMP algorithm by learning the optimal parameters from a large number of data. With the instantaneous CSI of the source-relay (SR) link, a detection network with SR channel (DNwSRC) is proposed applying deep unfolding approach when the ML detector is applied at the relay \cite{Jin2018Deep2}.
\vspace{-0.3em}
\subsection{Motivation and Contributions}

In this work, inspired by model-driven DL method, we propose a CG based DL network structure, namely learned CG network (LcgNet), for massive MIMO detection. The proposed network is constructed by unfolding the iterations of the CG algorithm, and each layer can be viewed as one iteration with some additional adjustable parameters. By following the prototype of LcgNet, two variants are proposed with scalar and vector network parameters, which are referred to as LcgNetS and LcgNetV, respectively. Furthermore,  in order to reduce the memory costs brought up by the
storage of the step-sizes, we present a novel quantized LcgNetV (QLcgNetV), where the step-sizes are smartly quantized by carefully designing a low-resolution nonuniform quantizer.

The main contributions of this work can be summarized as follows:

 {1) In the proposed LcgNet, we treat the step-sizes of the CG detector as learnable parameters. We show that the {detection} performance of LcgNetS is almost identical to that of the CG/LMMSE detector. Since the calculations of the step-sizes are simplified to some prestored parameters, the complexity of LcgNetS is much lower. Furthermore, significant performance gains over the CG/LMMSE detector  can be achieved by LcgNetV under various channel models.  The computational complexities of LcgNetV and LcgNetS are the same, but LcgNetV needs more memory space since more parameters are required to be stored.}

2) A novel QLcgNetV is proposed to save the memory costs resulted from the storage of the vector step-sizes. In QLcgNetV, a new nonuniform quantizer is employed and it is jointly optimized with LcgNetV to compress the network parameters. This quantizer is based on a specially designed soft
staircase function (referred to as the \emph{TanhSum} function), which  is constructed from a series of $\tanh(\cdot)$ functions. It is differentiable and has non-zero gradients everywhere. This appealing property allows us to integrate the proposed quantizer into the proposed network structure such that efficient training can be performed by backpropagation. We show that QLcgNetV can effectively reduce the required memory space with negligible detection performance loss. %{\color{blue}Furthermore, the proposed quantization scheme can be applied to many other fields to quantize the neural networks with less performance loss.}
%{\color{blue}
%Note that different from classical binary quantization methods \cite{Binaryconnect}, \cite{Binaryconnect1} which aim to quantize large-scale neural networks, the proposed quantizer can easily apply to many other fields where the number of the network parameters is limited. Compared with the fixed-format quantization schemes such as \cite{Gupta2015}, the proposed network can quantize the parameters into the data format of less bits with minor performance loss caused by quantization.}
%Note that different from other quantization methods which aims to quantize the large-scale neural networks \cite{Gupta2015}, \cite{Binaryconnect} and would cause severe performance loss when being employed by some relatively light networks, the proposed quantizer can easily apply to many other fields where the number of the network parameters is limited and the redundancy of the network is small.  }

3) Due to the fact that the number of learnable parameters in the proposed networks is very limited compared with some commonly known network structures, such as the fully-connected DNNs, our training process is relatively simple and easy to implement. {In our simulations,  the training can be completed  offline within 2 hours on a desktop with Intel (i3-6100) CPU running at 3.7GHz and 8GB RAM. Once trained, the proposed networks can be used to detect the transmit signal online through one forward pass, which takes about 0.003 seconds.}

In order to promote reproducible research, the Python codes for generating the main results in this work are available online at https://github.com/YiWei0129/LcgNet.
%In order to reduce the memory costs brought up by the storage of the step-sizes, we present a novel quantized LcgNetV (QLcgNetV), where some low-resolution nonuniform quantizers are integrated
%into the LcgNet to smartly quantize the aforementioned step-sizes. The specially designed low-resolution nonuniform quantizers, referred to as TanhSum, is consisting of a series of Tanh functions, thus is differentiable and has non-zero gradient values everywhere. This allows us to add the designed quantizer to the proposed network structure and exploit backpropagation with a stochastic gradient descent method to learn the optimal quantizer matched to the trained LcgNet. %During training, the soft staircase function is gradually annealed and eventually it converges to a discrete-valued hard one.
\vspace{-0.2cm}
\subsection{Organization of the Paper and Notations}
The rest of this paper is organized as follows.
Section II presents the system model and formulates the massive MIMO detection problem. Next, we introduce the main idea of the CG detector and present the proposed LcgNet in Section III.
%In section IV, we present the the complexity and memory cost analysis of the proposed networks.
Section IV provides the details of the proposed QLcgNetV. Numerical results are presented in Section V. Section VI concludes the paper. Finally, potential applications of our proposed LcgNet and some promising future research directions are given in Section VII.

\textit{Notation:} Scalars, vectors and matrices are respectively denoted by lower (upper) case, boldface lower case and boldface upper case letters. $x_n$ denotes the $n$-th entry of the vector $\bm{x}$. $\Re(\cdot)$ and $\Im(\cdot)$ denote the real and imaginary parts of their augments respectively; $||\cdot||$, $E[\cdot]$, $|\cdot|$, $(\cdot)^{-1}$ and $(\cdot)^H$ denote the L2 norm, expection, absolute, matrix inversion and conjugate transpose operations, respectively; ${\mathbb{C}}^{m\times n}$ (${\mathbb{R}}^{m\times n}$) denotes the space of $m \times n$ complex (real) matrices, and ${\mathbb{R}}^+$ denotes the space of positive real numbers. The symbol $\odot$ represents the Hadamard product. We define the complex normal distribution as $\mathcal{C}\mathcal{N}(\mu,{\sigma}^2)$ with mean $\mu$ and variance ${\sigma}^2$. Finally, $\textrm{sgn}(\cdot)$ is used to denote the signum function, where $\textrm{sgn}(x)=1$ when $x\ge0$, and $\textrm{sgn}(x)=-1$ otherwise.
\section{Preliminaries on MIMO detection}
\subsection{System Model}
  We consider a massive MIMO system with $N_t$ transmit antennas and $N_r$ receive antennas, where $N_t \le N_r$. Let $\bm{s} \in {\mathbb{A}^{{N_t}\times 1}}$ denote the unknown transmit signal vector, where $\mathbb{A}$ is the modulation alphabet in the complex valued constellations, and let $\bm{H}\in {\mathbb{C}}^{N_r \times N_t}$ denote the complex channel matrix. Then, the received signal vector $\bm{y} \in {\mathbb{C}}^{N_r \times 1}$ can be written as
  \begin{equation}\label{c}
    {\bm{y}} = {\bm{H}}{\bm{s}}+{\bm{n}},
  \end{equation}
  where ${\bm{n}}\sim\mathcal{C}\mathcal{N}(0,\sigma _n^2{\bm{I}}_{N_r})$ is the additive white Gaussian noise (AWGN).

 To facilitate the process of DL,  we avoid the handling of complex valued variables by employing an equivalent real-valued representation which is obtained by considering the real and imaginary parts separately. As a result, Eq. (\ref{c}) can be rewritten as follows:
  \begin{equation}
    {\bm{y}}_r = {\bm{H}_r}{\bm{s}}_r+{\bm{n}}_r,
  \end{equation}
  where
  \begin{equation}
  \begin{array}{l}
    {\bm{y}}_r =\left[ \begin{array}{l}
      \Re({\bm{y}})\\
      \Im({\bm{y}})\end{array}
    \right]\in {\mathbb{R}}^{2N_r\times 1},
   {\bm{s}}_r =\left[ \begin{array}{l}
      \Re({\bm{s}})\\
      \Im({\bm{s}})\end{array}
    \right]\in {\mathbb{R}}^{2N_t\times 1},\\
    \\
    {\bm{n}}_r =\left[ \begin{array}{l}
      \Re({\bm{n}})\\
      \Im({\bm{n}})\end{array}
    \right]\in {\mathbb{R}}^{2N_r\times 1},\\
    \\
     {\bm{H}}_r =\left[ \begin{array}{cc}
      \Re({\bm{H}}) & -\Im({\bm{H}})\\
      \Im({\bm{H}}) & \Re({\bm{H}})\end{array}
    \right]\in {\mathbb{R}}^{2N_r\times2N_t}.
  \end{array}
  \end{equation}
  \vspace{-1.5em}
  \subsection{Massive MIMO Detection}
%  The ML detector can achieve the optimal detection performance, measured by the minimum joint probability of error for detecting all the symbols simultaneously, however its computational complexity is exponential in $N_t$, which is prohibitive when the number of antennas is large.
{It is well-known that the ML detector can achieve the optimal detection performance, measured by the minimum joint probability of error for detecting all the symbols simultaneously. This optimal detector can be expressed as
  \begin{equation}\label{eq1}
 \hat{\bm{s}}_{\mathrm{ML}}=\arg \min _{\hat{\bm{s}} \in \mathbb{A}^{N_t\times1}}\|\bm{y}-\bm{H} \hat{\bm{s}}\|^{2},
 \end{equation}
 i.e., the message $\hat{\bm{s}}$ that
minimizes the distance between the received signal and the hypothesized noise-free message $\bm{H} \hat{\bm{s}}$ is selected.
 However, its computational complexity is exponential in $N_t$, which is prohibitive when the number of antennas is large. The SD detector can approach the performance of the ML detector with lower computational complexity by searching over fewer points. However, the complexity of the SD detector is related to the noise variance, search radius, $N_t$ and $N_r$, and its worst-case complexity is still exponential in $N_t$.
}

  Fortunately, the channel-hardening phenomenon in massive MIMO system offers new opportunities for signal detection. As shown by the Marcenko-Pastur law\cite{Mar1967DISTRIBUTION}, the singular values of $\bm{H}$ become less sensitive to the actual distributions of its i.i.d. entries and the diagonal entries of ${\bm{H}}^H\bm{H}$ will become increasingly larger in magnitude than the off-diagonal ones as the size of $\bm{H}$ increases. In other words, the channel becomes more and more deterministic with the increased number of antennas.
  %With the growth of the size of ${\bm{H}}^H\bm{H}$, the magnitudes of its diagonal entries converge to its dimension while the off-diagonal ones converge to $0$.
%\begin{figure}[t]
%\vspace{-0.4cm}
%\setlength{\abovecaptionskip}{-0.2cm}
%\setlength{\belowcaptionskip}{-0.5cm}
%\renewcommand{\captionfont}{\small}
%\centering
%\includegraphics[scale=.40]{CH.eps}
%\caption{An illustration of the channel hardening phenomenon.}
%\label{fig12}
%\normalsize
%\end{figure}
  With channel-hardening, simple linear detection algorithms are able to achieve good performance in massive MIMO systems, such as the ZF and LMMSE detectors. The main idea of these linear detectors is to obtain a preliminary estimation ${\hat{\bm{s}}}$ of the transmit symbol $\bm{s}$ by multiplying $\bm{y}$ with a receive filter $\bm{G}$, and then make decisions by mapping each element of ${\hat{\bm{s}}}$ into $\mathbb{A}$ according to the minimum distance criterion.
  The ZF and LMMSE detectors can be expressed as follows:
   \begin{equation}\label{Eq2}
    {\hat{\bm{s}}}_{\textrm{ZF}}={\bm{G}}_{\textrm{ZF}}\bm{y}=({\bm{H}}^H\bm{H})^{-1}{\bm{H}}^H\bm{y},
  \end{equation}
  \begin{equation}\label{Eq3}
    {\hat{\bm{s}}}_{\textrm{LMMSE}}={\bm{G}}_{\textrm{LMMSE}}\bm{y}=({\bm{H}}^H\bm{H}+{{\sigma}_n^2}{\bm{I}}_{N_t})^{-1}{\bm{H}}^H\bm{y},
  \end{equation}
   where ${\bm{I}}_{N_t}$ represents an $N_t\times N_t$ identity matrix. Compared with the ZF detector, the LMMSE detector takes the noise into consideration and therefore results in an improved performance. As can be seen from Eq. (\ref{Eq2}) and Eq. (\ref{Eq3}), both ZF and LMMSE detectors involve a matrix inversion operation. Since the dimension of the channel matrix in massive MIMO systems can be very large, their computational complexities will be considerable high.
\section{Learned CG Network}
In this section, we first review the CG algorithm for linear LMMSE detection\cite{7037382,7024849,7868573}, which is referred to as the CG detector in the following. It can iteratively achieve the performance of the LMMSE detector without matrix inversion. Then,  we present a detailed description of the proposed LcgNet and explain the intuition behind it. Finally, we provide the training strategy and complexity analysis of the proposed LcgNet.
%Finally, we provide the computational complexity analysis to show the advantage of the proposed network over conventional MMSE detection.
\vspace{-0.5em}
\subsection{The CG Detector}
  \begin{figure}[b]
\vspace{-0.7em}
\setlength{\belowcaptionskip}{-0.4cm}
\renewcommand{\captionfont}{\small}
\centering
\includegraphics[scale=.40]{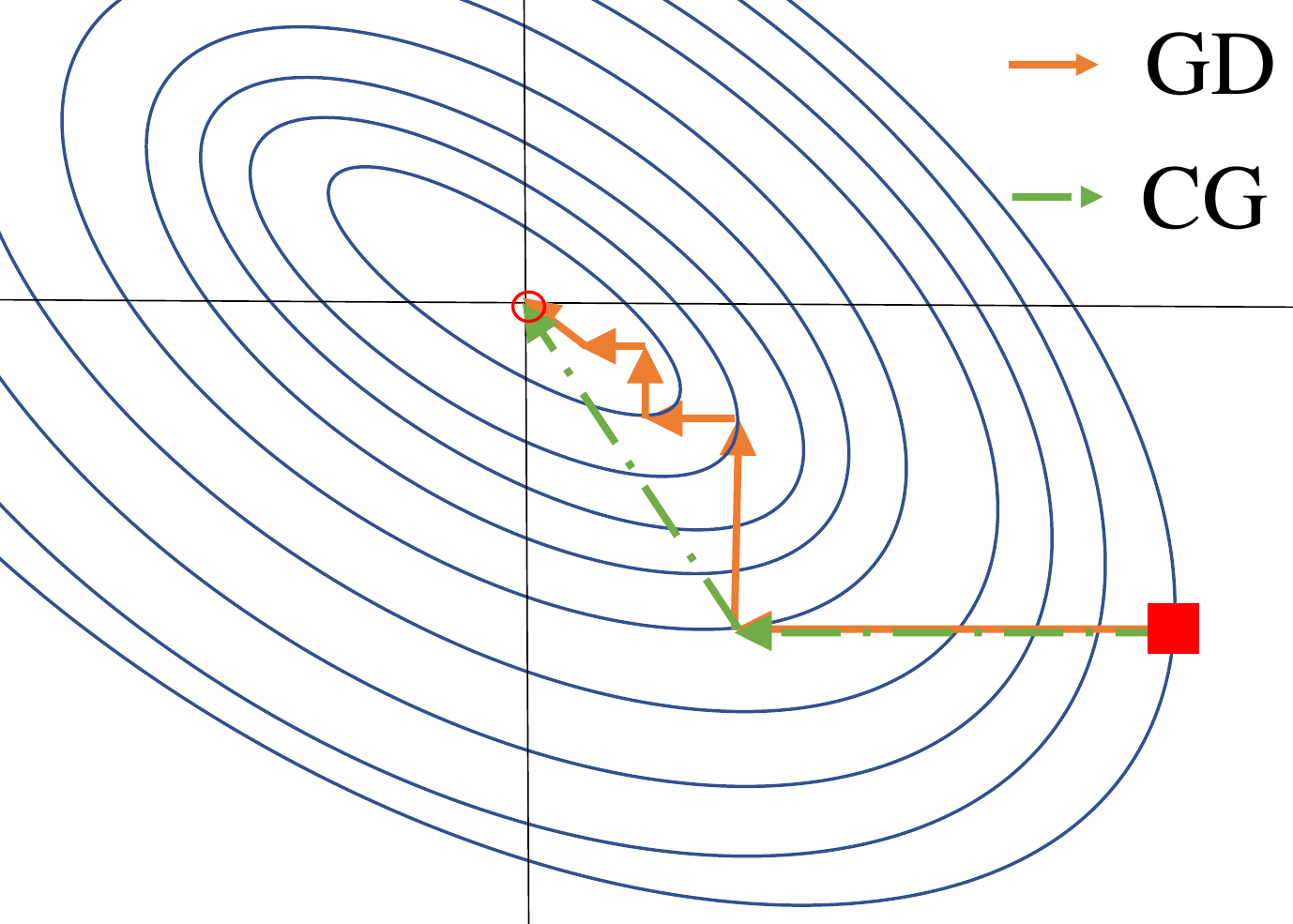}
\caption{The search procedures of the CG and GD algorithms.}
\label{fig1}
\normalsize
\end{figure}
Let us rewrite the LMMSE detector as follows:
\begin{equation}
\begin{aligned}
  {\hat{\bm{s}}}_{\textrm{LMMSE}}&=({\bm{H}}^H\bm{H}+{{\sigma}_n^2}{\bm{I}}_{N_t})^{-1}{\bm{H}}^H\bm{y}\\
  &={\bm{A}}_{\textrm{LMMSE}}^{-1}{\bm{b}}_{\textrm{LMMSE}},
\end{aligned}
\end{equation}
where ${\bm{A}}_{\textrm{LMMSE}}={\bm{H}}^H\bm{H}+{{\sigma}_n^2}{\bm{I}}_{N_t}$ and ${\bm{b}}_{\textrm{LMMSE}}={\bm{H}}^H\bm{y}$ denote the Hermitian positive definite LMMSE filtering matrix and matched-filter output vector, respectively. It can be seen that ${\hat{\bm{s}}}_{\textrm{LMMSE}}$ can be viewed as the solution of the linear equation ${\bm{A}}_{\textrm{LMMSE}}\bm{s}={\bm{b}}_{\textrm{LMMSE}}$.

  CG is an efficient iterative algorithm to solve this linear equation with low computational consumption. For ease of notation, the aforementioned linear equation can be generally rewritten as
  \begin{equation}
    \bm{As}=\bm{b}\label{a},
  \end{equation}
  where $\bm{A}\in {\mathbb{C}}^{K \times K}$ is a Hermitian positive definite matrix, $\bm{s}\in {{\mathbb{C}}^{K \times 1}}$ is the solution vector, and $\bm{b}\in {\mathbb{C}}^{K\times 1}$ is the measurement vector. Eq. (\ref{a}) can be equivalently transformed into the following quadratic optimization problem:
  \begin{equation}
    \mathop {\min}\limits_{\bm{s}} f(\bm{s})\triangleq\left(\frac{1}{2}{\bm{s}}^T\bm{As}-{\bm{b}}^T\bm{s}\right).
  \end{equation}
  Since $\bm{A}$ is symmetric and positive definite, the gradient of $f(\bm{s})$ at the optimal point $\hat{\bm{s}}$ would be zero, i.e., $f'(\hat{\bm{s}})=\bm{A}\hat{\bm{s}}-\bm{b}=\bm{0}$.
%Therefore,
%   \begin{equation}
%    \hat{\bm{x}}=\mathop {\arg\min}\limits_{\bm{x}}(f(\bm{x})=\frac{1}{2}{\bm{x}}^T\bm{Ax}-{\bm{b}}^T\bm{x})\Leftrightarrow \bm{A}\hat{\bm{x}}=\bm{b}.
%  \end{equation}
  Let $\mathcal{D}\triangleq\{{\bm{d}}^{(0)},{\bm{d}}^{(1)},\cdots,{\bm{d}}^{(K-1)}\}$ denote the conjugate direction set with respect to $\bm{A}$, i.e., ${\bm{d}}^{(i)H}\bm{A}{\bm{d}}^{(j)}=0, \forall i\ne j$.
%  Thus the conjugate direction of $i$ iteration can be derived as
%  \begin{equation}
%  {\bm{d}}^{(i)}={\bm{r}}^{(i)}-\mathop {\sum}\limits_{k<i}\frac{{{\bm{d}}^{(i)}}^H\bm{A}{\bm{r}}^{(i)}}{{{\bm{d}}^{(i)}}^H\bm{A}{\bm{d}}^{(i)}}{\bm{d}}^{(k)}.
%  \end{equation}
Then, we can minimize $f(\bm{s})$ in $K$ steps by successively minimizing it along $K$ individual conjugate directions in $\mathcal{D}$. By resorting to these conjugate directions, the CG algorithm usually exhibit faster convergence speed than conventional gradient descent algorithms, such as the steepest gradient descent algorithm (GD) \cite{Numerical}. The iterations of CG  can be described as
  \begin{equation}\label{b}
    {\hat{\bm{s}}}^{(i+1)}={\hat{\bm{s}}}^{(i)}+{\alpha}^{(i)}{\bm{d}}^{(i)},i=0,1,2,\cdots,
  \end{equation}
  where $i$ denotes the iteration index, and ${\alpha}^{(i)}$ is a scalar parameter which represents the step-size along direction ${\bm{d}}^{(i)}$. Furthermore, the residual ${\bm{r}}^{(i)}$ of the linear system (\ref{a}), which is also the descent direction of GD for $f({\hat{\bm{s}}}^{(i)})$, equals the negative gradient $-f'({\hat{\bm{s}}}^{(i)})$, i.e.,
  \begin{equation}\label{Eq8}
    {\bm{r}}^{(i)}=-f'({\hat{\bm{s}}}^{(i)})={\bm{r}}^{(i-1)}-{\alpha}^{(i-1)}\bm{A}{\bm{d}}^{(i-1)}.
  \end{equation}
  In the CG algorithm, each direction ${\bm{d}}^{(i)}$ is selected as a linear combination of the previous direction ${\bm{d}}^{(i-1)}$ and the negative gradient ${\bm{r}}^{(i)}$, i.e.,
  \begin{equation}\label{Eq9}
    {\bm{d}}^{(i)}={\bm{r}}^{(i)}+{\beta}^{(i-1)}{\bm{d}}^{(i-1)},
  \end{equation}
  where ${\beta}^{(i-1)}$ is a scalar parameter, serving as the step-size to update ${\bm{d}}^{(i)}$, and ${\bm{d}}^{(0)}$ is initialized as ${\bm{r}}^{(0)}$.

Fig. \ref{fig1} illustrates the search procedures of the CG and GD algorithms, where the ellipses denote the level faces of $f(\bm{s})$, the solid orange and dashed green arrows depict the descent
directions of GD (${\bm{r}}^{(i)}$) and CG (${\bm{d}}^{(i)}$), respectively. It can be observed that the search procedure of CG is not in zigzag shape, which shows that CG can achieve convergence with less iterations as compared with GD.

In addition, according to\cite{Numerical}, the step-sizes ${\alpha}^{(i)}$ and ${\beta}^{(i)}$ can be exactly calculated as follows:
\begin{equation}\label{d}
  {\alpha}^{(i)}=\frac{{\bm{r}}^{(i)H}{\bm{r}}^{(i)}}{{\bm{r}}^{(i)H}\bm{A}{\bm{d}}^{(i)}},
\end{equation}
\begin{equation}\label{e}
  {\beta}^{(i)}=\frac{{\bm{r}}^{(i+1)H}{\bm{r}}^{(i+1)}}{{\bm{r}}^{(i)H}{\bm{r}}^{(i)}}.
\end{equation}
%\subsection{CG-aided MMSE Detector}
\begin{algorithm}[t]
\setlength{\belowcaptionskip}{-0cm}
\setlength{\abovecaptionskip}{-0cm}
\caption{The CG detector} %??¡¤¡§¦Ì???¡Á?
\label{alg1}
\hspace*{0.02in} {\bf Input:} ${\bm{A}}$ and $\bm{b}$\\%??¡¤¡§¦Ì?¨º?¨¨?¡ê? \hspace*{0.02in}¨®?¨¤¡ä???????? ¡ê?¨ª?¨º¡À¨¤?¨®? \\ ??DD??DD
\hspace*{0.02in} {\bf Output:} Estimated transmit signal vector $\hat{\bm{s}}$% ??¡¤¡§¦Ì??¨¢1?¨º?3?
\begin{algorithmic}[1]
\State Initialization: $i=0$, ${\hat{\bm{s}}}^{(0)}=\bm{0}$, ${\hat{\bm{r}}}^{(0)}={\bm{b}}$, ${\hat{\bm{d}}}^{(0)}={\hat{\bm{r}}}^{(0)}$.% \State o¨®D¡ä¨°?¡ã?¨®???
\While{${\hat{\bm{r}}}^{(i)}\ne \bm{0}$}
    \State Update ${\alpha}^{(i)}$ according Eq. (\ref{d}),
    \State Update ${\hat{\bm{s}}}^{(i+1)}$ according to Eq. (\ref{b}),
    \State Update ${\bm{r}}^{(i+1)}$ according to Eq. (\ref{Eq8}),
    \State Update ${\beta}^{(i)}$ according to Eq. (\ref{e}),
    \State Update ${\bm{d}}^{(i+1)}$ according to Eq. (\ref{Eq9}),
    \State $i=i+1$.
\EndWhile
\State \Return $\hat{\bm{s}}={\hat{\bm{s}}}^{(i+1)}$.
\end{algorithmic}
\end{algorithm}
To summarize, the CG detector \cite{7024849} is listed in Algorithm \ref{alg1}.
\subsection{The Proposed LcgNet}
As one of the most popular and powerful schemes to build a model-driven DL network, unrolling a well-understood iterative algorithm is shown to outperform the baseline algorithm in many cases, such as the DetNet \cite{abs180507631} and the deep ADMM-net \cite{Yang2016Deep}.
%Inspired by this instructive idea, we design our LcgNet by mapping the iterations of Algorithm \ref{alg1} into a data flow graph.
Inspired by this instructive idea, we design our LcgNet by unfolding the iterations of Algorithm \ref{alg1} and transforming the step-sizes of each iteration into layer-dependent learnable parameters. The dimensions of the step-sizes can be augmented from scalars to vectors to further improve the detection performance. {For clarity, we refer to the networks with scalar and vector step-sizes as LcgNetS and LcgNetV, respectively. Compared with the CG detector, LcgNetV can achieve better detection accuracy with lower computational complexity, as will be shown in Section III-E and Section V.}

\begin{figure}
\vspace{-0.0cm}
\setlength{\belowcaptionskip}{-0.4cm}
\renewcommand{\captionfont}{\small}
\centering
\includegraphics[scale=.24]{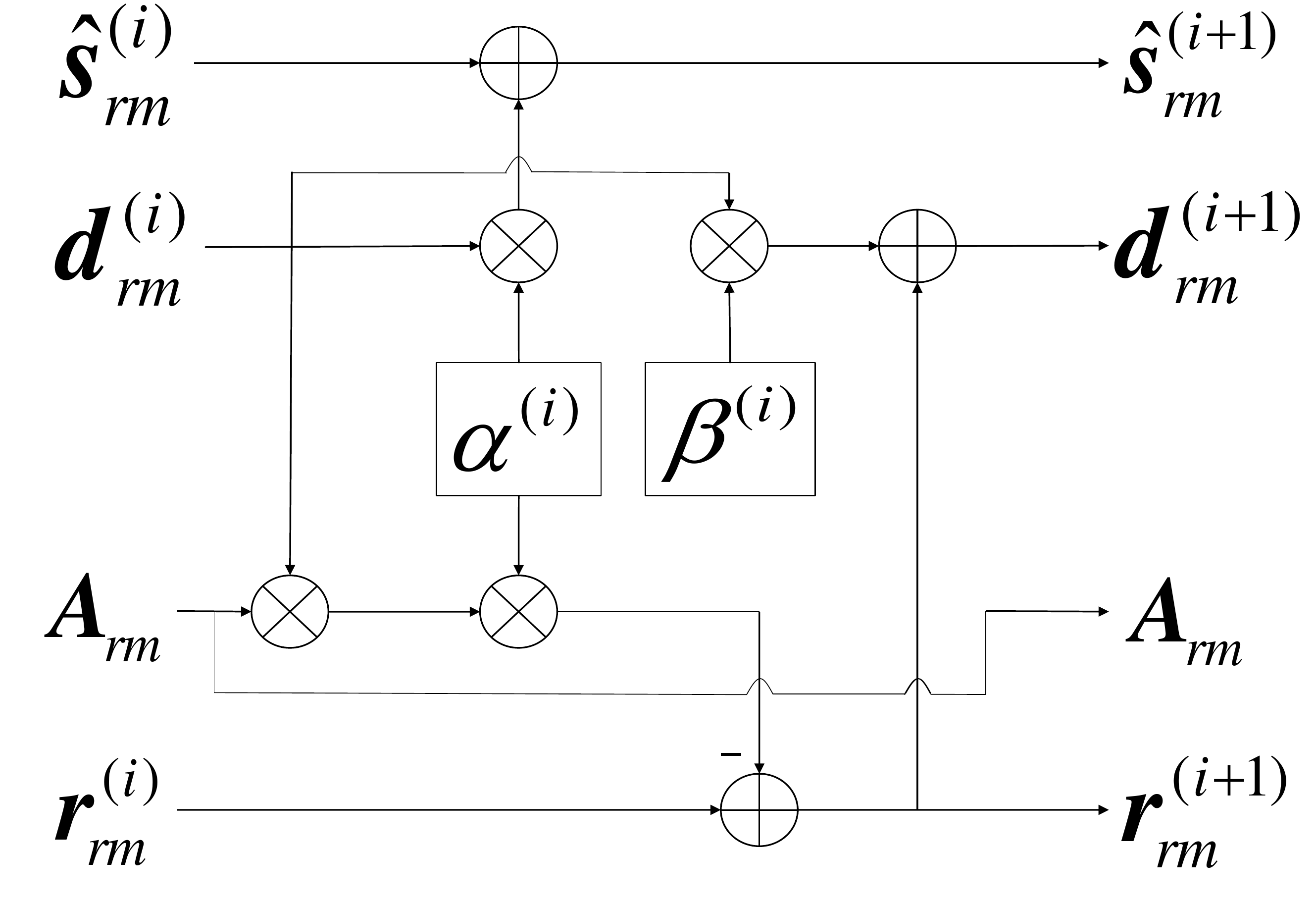}
\caption{The $i$-th layer of LcgNet with learnable parameters $\{{\alpha}_r^{(i)}, {\beta}_r^{(i)}\}$.}
\label{fig2}
\normalsize
\end{figure}

 {Let $\{({\bm{y}}_{rm},\bm{H}_{rm}),{\bm{s}}_{rm}\}^M_{m=1}$ denote the set of training samples with size $M$, where $({\bm{y}}_{rm},\bm{H}_{rm})$ and ${\bm{s}}_{rm}$ are the $m$-th feature and label, respectively.} Then, the proposed network is expected to accept {$({\bm{y}}_{rm},\bm{H}_{rm})$} as input and predict the label ${\bm{s}}_r$.
%  Based on the CG-aided MMSE detection algorithm, we unfold its iteration $i=1,2,\cdots, L$, resulting in a $L$-layer feed-forward neural network which is called LcgNet.
  Our deep {LcgNetS} is defined over a data flow graph based on the CG detector, which is shown in Fig. \ref{fig2}. The nodes in the graph correspond to different operations in CG, and the directed edges represent the data flows between these operations. The $i$-th iteration of the CG detector corresponds to the $i$-th layer of {LcgNetS}. Compared with the CG detector, whose step-sizes ${\alpha}^{(i)}$ and ${\beta}^{(i)}$ are calculated by (\ref{d}) and (\ref{e}) in the $i$-th iteration, we propose to introduce layer-dependent parameters ${\bm{\Theta}}^{(i)}=\{{\alpha}_r^{(i)}, {\beta}_r^{(i)}\}$ into the $i$-th layer of {LcgNetS} and learn these step-sizes from the training samples { $\{({\bm{y}}_{rm},\bm{H}_{rm}),{\bm{s}}_{rm}\}^M_{m=1}$} by minimizing the following mean square error (MSE) loss function:
  \begin{equation}\label{h}
  \begin{array}{l}
    {\mathcal{L}}^{(L)}( {\bm{\Theta}}^{(1)},\cdots,{\bm{\Theta}}^{(L)} )\\
     = \frac{1}{M}\sum\limits_{m = 1}^M {||{\bm{s}}_{rm} - {\hat{\bm{s}}}_r^{(L)}( {\bm{y}}_{rm},\bm{H}_{rm};{\bm{\Theta}}^{(1)},\cdots,{\bm{\Theta}}^{(L)})||^2}.
  \end{array}
  \end{equation}
In (\ref{h}), $L$ denotes the number of layers, and {${\hat{\bm{s}}}_r^{(L)}({\bm{y}}_{rm},\bm{H}_{rm};{\bm{\Theta}}^{(1)},\cdots,{\bm{\Theta}}^{(L)})$} is the output of {LcgNetS} with ${\bm{y}}_{rm}$ as inputs and $\{{\bm{\Theta}}^{(1)},\cdots,{\bm{\Theta}}^{(L)}\}$ as parameters. As illustrated in Fig. \ref{fig2}, the $i$-th layer of {LcgNetS} can be represented by
  \begin{subequations}\label{16}
  \begin{equation}
  {\hat{\bm{s}}}_{rm}^{(i+1)}={\hat{\bm{s}}}_{rm}^{(i)}+{\alpha}_r^{(i)}{\bm{d}}_{rm}^{(i)},\label{16-1}
  \end{equation}
    \begin{equation}
  {\bm{r}}_{rm}^{(i+1)}={\bm{r}}_{rm}^{(i)}-{\alpha}_r^{(i)}{\bm{A}}_{rm}{\bm{d}}_{rm}^{(i)},
  \end{equation}
  \begin{equation}
  {\bm{d}}_{rm}^{(i+1)}={\bm{r}}_{rm}^{(i+1)}+{\beta}_r^{(i)}{\bm{d}}_{rm}^{(i)},\label{16-3}
  \end{equation}
  \end{subequations}
 where ${\hat{\bm{s}}}_{rm}^{(0)}=\bm{0}$, ${{\bm{r}}}_{rm}^{(0)}={\bm{b}}_{rm}$ and $ {{\bm{d}}}_{rm}^{(0)}={{\bm{r}}}_{rm}^{(0)}$ denote the first-layer inputs,  ${\bm{A}}_{rm}={\bm{H}}_{rm}^T{\bm{H}}_{rm}+{{\sigma}_n^2}{\bm{I}}_{2N_t}$ and ${\bm{b}}_{rm}={\bm{H}}_{rm}^T{\bm{y}}_{rm}$. It can be seen that in {LcgNetS}, the calculations of ${\alpha}_r^{(i)}$ and ${\beta}_r^{(i)}$, which originally involve matrix-vector multiplication and division operations, are replaced by some prestored parameters which are fixed during online detections. Note that this can effectively reduce the computational complexities, only at the expense of some additional memory costs.

\begin{figure*}[b]
\vspace{-0.3cm}
\setlength{\belowcaptionskip}{-0.4cm}
\renewcommand{\captionfont}{\small}
\centering
\includegraphics[scale=.53]{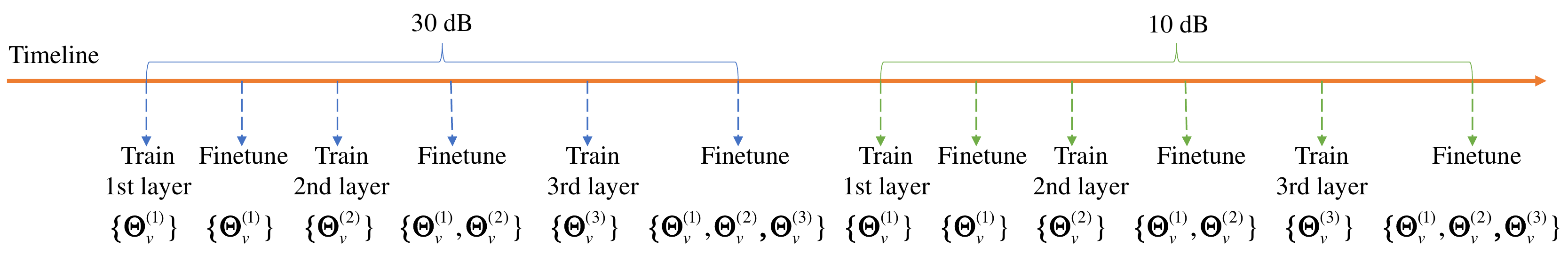}
\caption{{Training process of a $3$-layer LcgNetV (SNR = [30, 10] $\textrm{dB}$).}}
\label{table}
\normalsize
\end{figure*}
%\begin{table*}[b]
%\setlength{\abovecaptionskip}{0.3cm}
%\setlength{\belowcaptionskip}{0cm}
%\caption{Training process of a $3$-layer LcgNetV (SNR = [30, 10] $\textrm{dB}$)}
%\centering
%\begin{tabular}{|c|c|c|c|c|c|c|c|c|c|c|c|c|}
%\hline
%\multirow{3}*{${\bm{\Theta}}_v^{(i)}$} & \multicolumn{6}{c|}{30 dB} &  \multicolumn{6}{c|}{10 dB}\\
%\cline{2-13}
% &\multicolumn{2}{c|} {1-th layer} & \multicolumn{2}{c|}{2-th layer} & \multicolumn{2}{c|}{3-th layer} &  \multicolumn{2}{c|}{1-th layer} & \multicolumn{2}{c|}{2-th layer} & \multicolumn{2}{c|}{3-th layer} \\
%\cline{2-13}
%& Train & Finetune & Train & Finetune & Train & Finetune & Train & Finetune & Train & Finetune & Train & Finetune \\
% \hline
%${\bm{\Theta}}_v^{(1)}$ &   $\surd$ & { $\surd$} & &  $\surd$ & & { $\surd$} & { $\surd$} & { $\surd$} & & {$\surd$} & & { $\surd$}\\
%\hline
%${\bm{\Theta}}_v^{(2)}$ &   &  & {$\surd$} & { $\surd$} & & { $\surd$} &   &  & { $\surd$} & { $\surd$} & & { $\surd$}  \\
%\hline
%${\bm{\Theta}}_v^{(3)}$ &   &  &  &  &{ $\surd$} & { $\surd$}  &   &  &  &  &{ $\surd$} & { $\surd$}  \\
%\hline
%\end{tabular}
%\label{Tab3}
%\end{table*}
   Moreover, we can further improve the performance of {LcgNetS} by lifting the scalar parameter ${\bm{\Theta}}^{(i)}$ to a higher dimension, e.g., using vector step-sizes ${\bm{\Theta}}^{(i)}_v = \{{{\bm{\alpha}}_r^{(i)}}\in {\mathbb{R}}^{2N_t\times 1}, {{\bm{\beta}}_r^{(i)}}\in {\mathbb{R}}^{2N_t\times 1}\}$\footnote{We have also investigated the case where matrix step-sizes are employed, however, only minor performance gain is observed. Therefore, in this work, we only consider the use of vector step-sizes.}. By following this idea, we are able to learn the appropriate values of these high dimensional step-sizes and this may have the potential to even outperform the original CG detector. For LcgNetV, similar to (\ref{16}), the operations involved in the $i$-th layer can be expressed as
  \begin{subequations}\label{20}
     \begin{equation}
  {\hat{\bm{s}}}_{rm}^{(i+1)}={\hat{\bm{s}}}_{rm}^{(i)}+{\bm{\alpha}}_r^{(i)}\odot{\bm{d}}_{rm}^{(i)},\label{17-1}
  \end{equation}
    \begin{equation}
  {\bm{r}}_{rm}^{(i+1)}={\bm{r}}_{rm}^{(i)}-{\bm{\alpha}}_r^{(i)}\odot{\bm{A}}_{rm}{\bm{d}}_{rm}^{(i)},
  \end{equation}
  \begin{equation}
  {\bm{d}}_{rm}^{(i+1)}={\bm{r}}_{rm}^{(i+1)}+{\bm{\beta}}_r^{(i)}\odot{\bm{d}}_{rm}^{(i)}.\label{17-3}
  \end{equation}
  \end{subequations}
Note that the computational complexity of (\ref{20}) is almost identical to that of (\ref{16}) in LcgNetS, however, in this case, more parameters are needed to be stored.
\subsection{The Intuition behind the Proposed LcgNetV}
In this subsection, we give the main motivation behind the proposed LcgNetV, which intuitively {explains} the reason why such an approach works.

For the sake of notation simplicity, we consider a $K$-dimensional vector space $\mathcal{V}$.
  Initialized at ${\hat{\bm{s}}}^{(0)}$, any gradient descent-type algorithms choose a direction ${\bm{d}}^{(i)}$ and then search along this direction with a proper step-size $\sigma^{(i)}$ for a new iteration in order to achieve a smaller distance to the optimal solution ${\bm{s}}^*$. Suppose that we can find the optimal solution after $I$ iterations, i.e.,
  \begin{equation}\label{17}
    {\bm{s}}^* = {\hat{\bm{s}}}^{(0)}+{\sigma}^{(0)}{\bm{d}}^{(0)}+\cdots+{\sigma}^{(I-1)}{\bm{d}}^{(I-1)}.
  \end{equation}
  %where $\{{\sigma}^{(i)}\}_{i=0}^{I-1}$ denotes the set of step-sizes. Then, we can express the difference between ${\hat{\bm{s}}}^{(0)}$ and ${\bm{s}}^*$ as follows:
%    \begin{equation}\label{f}
%    {\bm{s}}^* - {\hat{\bm{s}}}^{(0)}= {\sigma}^{(0)}{\bm{d}}^{(0)}+\cdots+{\sigma}^{(I-1)}{\bm{d}}^{(I-1)}.
%  \end{equation}
  One can view the above expression as gradient descent  trying to find a set of vectors and the corresponding weights (step-sizes) whose weighted linear combination is equivalent to ${\bm{s}}^* - {\hat{\bm{s}}}^{(0)}$. Note that an arbitrary vector in space $\mathcal{V}$ can be expressed as a linear combination of the basis vectors, and any set of $K$ linearly independent vectors in $\mathcal{V}$ is automatically a basis for this space.

The CG algorithm tries to find a set of conjugate directions, which are mutually independent \cite{Numerical}, such that the number of iterations can be restricted to the space dimension.
{In the proposed LcgNetS, since ${\sigma}^{(i)}$ are fixed during online signal detection, ${\bm{d}}^{(i)}$ can be obtained according to \eqref{16-1}-\eqref{16-3}, which only contain linear operations (addition or scalar-vector multiplication) and hence LcgNetS is in general a linear detector.
  For LcgNetV, however, (\ref{17}) is transformed into
  \begin{equation}\label{g}
    {\bm{s}}^* - {\hat{\bm{s}}}^{(0)}={\bm{\gamma}}^{(0)}\odot{\bm{d}}^{(0)}+\cdots+{\bm{\gamma}}^{(I-1)}\odot{\bm{d}}^{(I-1)},
  \end{equation}
  where step-sizes $\{{\bm{\gamma}}^{(i)}\}_{i=0}^{I-1}$ are in vector form.
  {Since the non-linear Hadamard product operation is introduced, LcgNetV is essentially a non-linear detector.} Therefore, by resorting to the power of DL, we can train the high-dimensional parameters $\{{\bm{\gamma}}^{(i)}\}_{i=0}^{I-1}$ which can be applied to multiple scenarios with arbitrary ${\bm{s}}^{*}$, and the resulting LcgNetV can achieve better performance than linear detectors. Generally, other networks which can perform non-linear operations, such as the fully-connected network, etc., also have ability to outperform linear detectors. However, the performance gains of these networks usually come at the cost of  high computational complexity and huge number of parameters. The proposed LcgNetV can be viewed as a light-weight nonlinear detector, which can achieve a good  tradeoff between network complexity and detection performance.}
%, i.e., $I=K$.
%  In our LcgNetV, (\ref{17}) is transformed into
%  \begin{equation}\label{g}
%    {\bm{s}}^* - {\hat{\bm{s}}}^{(0)}={\bm{\sigma}}^{(0)}\odot{\bm{d}}^{(0)}+\cdots+{\bm{\sigma}}^{(I-1)}\odot{\bm{d}}^{(I-1)},
%  \end{equation}
%  where the step-sizes $\{{\bm{\sigma}}^{(i)}\}_{i=0}^{I-1}$ are in vector form.
%  Consider a single problem instance, it can be readily seen that with a given ${\bm{s}}^{*}=[s_1^{*},s_2^{*},\cdots,s_K^{*}]^T$ and a prechosen ${\bm{d}}^{(0)}=[d_1^{(0)},d_2^{(0)},\cdots,d_K^{(0)}]^T$, where ${\{d_j^{(0)}\}_{j=1}^K}\ne0$, we can always find a vector step-size ${\bm{\sigma}}^{(0)}=[s_1^*/d_1^{(0)},s_2^*/d_2^{(0)},\cdots,s_K^*/d_K^{(0)}]^T$ such that Eq. (\ref{g}) can be satisfied. However, it is impossible to satisfy Eq. (\ref{17}) with only one iteration when the step-sizes are in scalar form. Therefore, by resorting to the power of DL, LcgNetV is expected to learn these high-dimensional parameters that can be applied to multiple scenarios with arbitrary ${\bm{s}}^{*}$. Simulation results corroborates that LcgNetV can indeed reduce the number of iterations and outperform the conventional CG detector, which will be elaborated in Section V.
\subsection{Training Details}
\begin{table*}[b]
\vspace{-0.2em}
\setlength{\abovecaptionskip}{-0.0cm}
\setlength{\belowcaptionskip}{-0.0cm}
\caption{Complexity analysis of the considered detectors}
\begin{center}
\begin{tabular}{lc|cc|cc}
\hline
\multicolumn{2}{c|}{The LMMSE detector} & \multicolumn{2}{c|}{The CG detector} & \multicolumn{2}{c}{LcgNet}\\
\hline
Calculating ${\bm{A}}_{\textrm{LMMSE}}^{-1}$ & $\mathcal{O}(8N_t^3)$ & Updating ${\alpha}^{(i)}$ & $\mathcal{O}(4N_t^2+4N_t+4)+1$ $\textrm{RDiv}$ & & \\
Calculating ${\bm{A}}_{\textrm{LMMSE}}^{-1}\times{\bm{b}}_{\textrm{LMMSE}}$ & $\mathcal{O}(4N_t^2)$ & Updating ${\beta}^{(i)}$ & $\mathcal{O}(4N_t+4)+1$ $\textrm{RDiv}$ & & \\
 & & Updating ${\hat{\bm{s}}}^{(i)}$ & $\mathcal{O}(2N_t)$ & Updating ${\hat{\bm{s}}}^{(i)}$ & $\mathcal{O}(2N_t)$ \\
   & & Updating ${\bm{r}}^{(i)}$ & $\mathcal{O}(4N_t^2+2N_t)$ & Updating ${\bm{r}}^{(i)}$ & $\mathcal{O}(4N_t^2+2N_t)$ \\
 & & Updating ${\bm{d}}^{(i)}$ & $\mathcal{O}(2N_t)$ & Updating ${\bm{d}}^{(i)}$ & $\mathcal{O}(2N_t)$ \\
\hline
\multicolumn{2}{c|}{$\mathcal{O}(8N_t^3+4N_t^2)$} & \multicolumn{2}{c|}{$\mathcal{O}(L(8N_t^2+14N_t+8))+2L$ $\textrm{RDiv}$} & \multicolumn{2}{c}{$\mathcal{O}(L(4N_t^2+6N_t))$}\\
\hline
\end{tabular}
\label{Tab1}
\end{center}
\end{table*}
The proposed networks are implemented in Python using the TensorFlow library with the Adam optimizer \cite{ADAM}.

\emph{Training/testing data}: Since the proposed networks are expected to work at various signal-to-noise ratio (SNR) levels, we construct the training data set by randomly generating the training samples {$\{({\bm{y}}_{rm},\bm{H}_{rm}),{\bm{s}}_{rm}\}^M_{m=1}$} based on (\ref{c}) with different levels of channel noise. The transmit symbols in $\bm{s}$ are from {the constellation points of a specific modulation scheme (e.g., BPSK, QPSK or 16QAM)} and $\bm{H}$ is randomly generated according to some specific channel models, {such as the Rayleigh fading channel, the spatial correlated channel \cite{951380} and the TDL-A MIMO channel \cite{3GPP38}}. The training and testing data sets contain $10^5$ and $10^4$ samples, respectively.

\emph{Training process}: We first train the proposed networks using training samples with high SNR, e.g., SNR = 30 $\textrm{dB}$, in order to learn the intrinsic structure of the detection problem. Then, we employ the samples with lower SNRs in the subsequent training process for the purpose of reducing the influence caused by noise. More specifically, in order to gradually improve the robustness of the proposed networks against channel noise, the SNR levels of the training samples are chosen from the range $[25,20,15,10,5,0]$ $\textrm{dB}$ successively. From the beginning, the step-sizes $\{{\bm{\Theta}}^{(i)}\}_{i=1}^{L}$ and $\{{\bm{\Theta}}_v^{(i)}\}_{i=1}^L$ are initialized as zeros. For each SNR level, we train the proposed network with the same strategy, which is described in details as follows.

%The step-sizes ${\bm{\Theta}}$ and ${\bm{\Theta}}_v$ are trained by beginning with a single-layer network and then gradually increasing the network size one layer at a time. More specifically, when training the LcgNetS (LcgNetV) with $l$ layers, we first train the parameters $\{{\alpha}_r^{(l)}, {\beta}_r^{(l)}\}$ ($\{{\bm{\alpha}}_r^{(l)}, {\bm{\beta}}_r^{(l)}\}$) in the current layer with $\{{\alpha}_r^{(i)}, {\beta}_r^{(i)}\}^{l-1}_{i=1}$ ($\{{\bm{\alpha}}_r^{(i)}, {\bm{\beta}}_r^{(i)}\}^{l-1}_{i=1}$) fixed, and the learning rate is set to be 0.001. After that, we finetune all the parameters $\{{\alpha}_r^{(i)}, {\beta}_r^{(i)}\}^l_{i=1}$ ($\{{\bm{\alpha}}_r^{(i)}, {\bm{\beta}}_r^{(i)}\}^l_{i=1}$) by using a decaying learning rate which is set to be 0.0005 initially and then reduce it by half in every epoch.
{\
%In order to  acquire more information
%about the network with different numbers of layers, the proposed network is trained in a layer-by-layer learning strategy, i.e., we begin to train with a
%single-layer network and then gradually increase the network size one layer at a time. If we train the whole neural network with $L$ layers in an end-to-end manner, only the performance of an $L$-layer network model can be observed. In contrast, by employing the proposed layer-by-layer learning strategy, the performance of network with various numbers of layers ($\{1,2,\cdots ,L\}$) can be observed, thus we can strike a good balance between the detection performance and detection complexity.  Specifically, the training process of the $l$-th layer can be divided into two steps: (1) the parameters of the current layer (i.e., the l-th layer) are trained with the parameters of the preceding $l-1$ layers fixed, and (2) the parameters contained in these $l$ layers are further finetuned. In the first step,  the learning rate is set to be a relatively large value, i.e., 0.001, and in the second step, we use a decaying learning rate which is set to be 0.0005 initially and then reduce it by half in every epoch to finetune the parameters $\{{\alpha}_r^{(i)}, {\beta}_r^{(i)}\}^l_{i=1}$ ($\{{\bm{\alpha}}_r^{(i)}, {\bm{\beta}}_r^{(i)}\}^l_{i=1}$).
In order to  extract more information
about the network when different numbers of layers are employed, the proposed network is trained in a layer-by-layer way, i.e., we begin to train a
single-layer network and then gradually increase the network size one layer at a time. Note that if we train the whole  network with $L$ layers in an end-to-end manner, then only the performance of an $L$-layer network can be observed. Instead, by employing the proposed layer-by-layer learning strategy, the performance of the network with various numbers of layers ($\{1,2,\cdots ,L\}$) can be observed, thus we can strike a good balance between detection performance and detection complexity  by employing an appropriate number of layers.  Specifically, the training process of the $l$-th layer can be divided into two steps: (1) the parameters of the current layer (i.e., the l-th layer) are trained with the parameters of the preceding $l-1$ layers fixed, and (2) {the parameters of these $l$ layers} are further finetuned. In the first step,  the learning rate is set to be a relatively large value, i.e., 0.001, and in the second step, we use a decaying learning rate which is set to be 0.0005 initially and then reduce it by half in every epoch to finetune the parameters $\{{\bm{\Theta}}^{(i)}\}_{i=1}^{l}$ ($\{{\bm{\Theta}}_v^{(i)}\}_{i=1}^l$).}
%In Order To  Extract More Information
%About The Network With Different Numbers Of Layers, The Proposed Network Is Trained In A Layer-By-Layer Way, I.E., We Begin To Train With A
%Single-Layer Network And Then Gradually Increase The Network Size One Layer At A Time. Note That If We Train The Whole  Network With $L$ Layers In An End-To-End Manner, Then Only The Performance Of An $L$-Layer Network Model Can Be Observed. Instead, By Employing The Proposed Layer-By-Layer Learning Strategy, The Performance Of Network With Various Numbers Of Layers ($\{1,2,\Cdots ,L\}$) Can Be Observed, Thus We Can Strike A Good Balance Between The Detection Performance And Detection Complexity  By Employing An Appropriate Number Of Layers.  Specifically, The Training Process Of The $L$-Th Layer Can Be Divided Into Two Steps: (1) The Parameters Of The Current Layer (I.E., The L-Th Layer) Are Trained With The Parameters Of The Preceding $L-1$ Layers Fixed, And (2) {\Color{Blue}The Parameters Of These $L$ Layers} Are Further Finetuned. In The First Step,  The Learning Rate Is Set To Be A Relatively Large Value, I.E., 0.001, And In The Second Step, We Use A Decaying Learning Rate Which Is Set To Be 0.0005 Initially And Then Reduce It By Half In Every Epoch To Finetune The Parameters $\{{\Alpha}_R^{(I)}, {\Beta}_R^{(I)}\}^L_{I=1}$ ($\{{\Bm{\Alpha}}_R^{(I)}, {\Bm{\Beta}}_R^{(I)}\}^L_{I=1}$). }
All these training processes are terminated
when the average validation normalized MSE (NMSE) stops decreasing.
For clarity, we illustrate the training process of a 3-layer LcgNetV in Fig. \ref{table} as a toy example, where the SNR levels of the training samples are chosen from $[30,10]$ $\textrm{dB}$ successively.  %As shown in Table \ref{Tab3}, we mark the parameters required training of every training steps.

 Furthermore, since the number of learnable parameters that are required to be optimized is limited, the training process of our proposed networks is relatively easy and simple, and it took about several hours on a standard Intel i3-6100 processor. The training time would increase with the expansion of the MIMO scale and the dimension of the learnable parameters.

\subsection{Complexity and Memory Cost Analysis}
This subsection provides the computational complexity analysis of the LMMSE detector, the CG detector with $L$ iterations, the $L$-layer LcgNetS and LcgNetV. The comparison is based on the number of real multiplications needed in one online detection process. In our analysis, one complex multiplication equals four real multiplications and one complex division equals four real multiplications plus one real division (RDiv). %Besides, the number of complex divisions (CDivs) is also taken into consideration.

{

Note that one scalar-vector multiplication, i.e., $\epsilon \bm{v}$, $\bm{v}\in {\mathbb{R}}^{c\times 1}$, $\epsilon\in \mathbb{R}$, contains $c$ real multiplications and so does one vector-vector multiplication, i.e., ${\bm{v}}^T \bm{v}$, one vector-matrix multiplication, i.e., ${\bm{v}}^T \bm{V}$ or ${\bm{V}}^T {\bm{v}}$, $\bm{V}\in {\mathbb{R}}^{c\times b}$,  needs $cb$ real multiplications, and the computational complexity of one matrix inversion operation, i.e., ${\bm{S}}^{-1}$, $\bm{S}\in {\mathbb{R}}^{c\times c}$, is on the order of  $\mathcal{O}(c^3)$ \cite{6954512}. Suppose that ${\bm{A}}_{\textrm{LMMSE}}$ and ${\bm{b}}_{\textrm{LMMSE}}$ are calculated in advance for all the considered detectors, then the computational complexity of the LMMSE detector involves: $1)$ the inversion of ${\bm{A}}_{\textrm{LMMSE}}\in {\mathbb{R}}^{2N_t\times 2N_t}$, which is on the order of $\mathcal{O}(8N_t^3)$; $2)$ the vector-matrix multiplication ${\bm{A}}_{\textrm{LMMSE}}^{-1}{\bm{b}}_{\textrm{LMMSE}}$, which requires $\mathcal{O}(4N_t^2)$ complexity. The complexity of the CG detector in each iteration includes: $1)$ calculating ${\alpha}^{(i)}$ in \eqref{d}, which contains one vector-matrix multiplication (${\bm{r}}^{(i)H}\bm{A}$, $\bm{A} \in {\mathbb{R}}^{2N_t \times 2N_t}$), two vector-vector multiplications (${\bm{r}}^{(i)H} {\bm{r}}^{(i)}$ and ${\bm{r}}^{(i)H}\bm{A}\times {\bm{d}}^{(i)}$, ${\bm{r}}^{(i)}\in{\mathbb{R}}^{2N_t\times 1}$, ${\bm{d}}^{(i)}\in {\mathbb{R}}^{2N_t\times 1}$) and one complex division $\big(({{\bm{r}}^{(i)H}{\bm{r}}^{(i)}})/({{\bm{r}}^{(i)H}\bm{A}{\bm{d}}^{(i)}})\big)$, thus the computational complexity is $\mathcal{O}(4N_t^2+4N_t+4)+1$ $\textrm{RDiv}$; $2)$ calculating ${\beta}^{(i)}$ in \eqref{e}, which includes two vector-vector multiplications (${\bm{r}}^{(i+1)H}{\bm{r}}^{(i+1)}$ and ${\bm{r}}^{(i)H}{\bm{r}}^{(i)}$) and one complex division $\big(({{\bm{r}}^{(i+1)H}{\bm{r}}^{(i+1)}})/({\bm{r}}^{(i)H}{{\bm{r}}^{(i)}})\big)$, thus its complexity can be expressed as $\mathcal{O}(4N_t+4)+1$ $\textrm{Rdiv}$; $3)$ updating $\{{\hat{\bm{s}}}^{(i)},{\bm{d}}^{(i)},{\bm{r}}^{(i)}\}$, which requires one vector-matrix multiplication (${\bm{A}}{\bm{d}}^{(i)}$) and three scalar-vector multiplications ( ${\alpha}^{(i)}{\bm{d}}^{(i)}$, ${\beta}_r^{(i)}{\bm{d}}^{(i)}$ and ${\alpha}_r^{(i)} {\bm{A}}{\bm{d}}^{(i)}$), hence the complexity is given by $\mathcal{O}(4N_t^2+6N_t)$.}
The LcgNetS and LcgNetV exhibit the same computational complexity, which can be easily obtained by removing the calculations of ${\alpha}_r^{(i)}$ and ${\beta}_r^{(i)}$ in the CG detector. See TABLE \ref{Tab1} for a summary. {
We can see that the computational complexity of LcgNet is similar to that of the GS-based detector \cite{6954512}, which can be expressed as $\mathcal{O}((L_{\textrm{GS}} + 1)4N_t^2 + 8N_t)$ with $L_{\textrm{GS}}$ denoting the iteration number. However, the detection performance of the proposed LcgNetV is better  since the GS-based detector cannot outperform the LMMSE detector.}

Furthermore, we investigate the memory costs of the CG detector, LcgNetS and LcgNetV, where we assume that $B$ bits are required to store one real number. For the CG detector, only $\hat{\bm{s}}^{(i)},{\bm{d}}^{(i)}$, ${\bm{r}}^{(i)}$ and ${\bm{A}}_{\textrm{LMMSE}}$ are needed to be stored for the next iteration, whose memory cost is $(6N_t+4N_t^2)B$ bits. In the online detection process of the proposed networks, other than $\hat{\bm{s}}_{rm}^{(i)},{\bm{d}}_{rm}^{(i)}$, ${\bm{r}}_{rm}^{(i)}$ and ${\bm{A}}_{rm}$, the step-sizes $\bm{\Theta}$ and ${\bm{\Theta}}_v$ are additional parameters that are required to be prestored in LcgNetS and LcgNetV, thus their memory cost can be expressed as $(6N_t+4N_t^2+4L)B$ bits and $(6N_t+4N_t^2+4LN_t)B$ bits, respectively.

\section{The proposed quantized LcgNetV}
  In the previous section, we can see that the advantage of the CG detector is its low memory costs, i.e., only $\hat{\bm{s}}^{(i)},{\bm{d}}^{(i)}$, ${\bm{r}}^{(i)}$ and ${\bm{A}}_{\textrm{LMMSE}}$ are needed to be stored for the next iteration. Nevertheless, in our proposed LcgNet, we need to store all the step-sizes (scalar step-sizes ${\bm{\Theta}}$ in LcgNetS and vector step-sizes ${\bm{\Theta}}_v$ in LcgNetV) since they are fixed for all problem instances. In order to reduce this memory cost, in this section, we further present a novel QLcgNetV, which is based on a specially designed soft staircase function with adjustable parameters. In the following, we will first introduce the conventional hard quantizer, and then by combining the trained LcgNetV and the proposed soft staircase function, we present the QLcgNetV, where these two important components are jointly trained. By properly optimizing the learnable parameters in the proposed soft staircase function, nonuniform quantization can be achieved with higher compression efficiency.
  \subsection{Hard Quantizer}
 \newcounter{TempEqCnt1}
\setcounter{TempEqCnt1}{\value{equation}}
\setcounter{equation}{24}
  \begin{figure*}[hb]
  \hrulefill
   \begin{equation} \label{j}
         {\mathcal{L}}^{(L)}_{\textrm{Qs}}( \bm{\Phi}) \triangleq \frac{1}{M}\sum\limits_{m = 1}^M {||{\bm{s}}_{rm} - {\hat{\bm{s}}}_r^{(L)}( {\bm{y}}_{rm},\bm{H}_{rm};Q_s({\bm{\Theta}}^{(1)}_v;\bm{\Phi},\{\sigma,l,G_{{b}}\}),\cdots,Q_s({\bm{\Theta}}^{(L)}_v;\bm{\Phi},\{\sigma,l,G_{{b}}\}))||^2}
    \end{equation}
   \hrulefill
\end{figure*}
\setcounter{TempEqCnt1}{\value{equation}}
\setcounter{equation}{18}
  A quantizer can be seen as a real-valued function $Q(x)$ which maps $x\in\mathbb{R}$ to a finite set $\mathcal{G}\subset\mathbb{R}$. For a vector $\bm{x}$, we define $Q(\bm{x})\buildrel \Delta \over = (Q(x_1),Q(x_2),\cdots,Q(x_n))^T$.
    The staircase function is one of the most commonly used quantizers (also referred to as the hard quantization function), which can be expressed as
    \begin{equation}\label{Eq25}
      Q_h(x)= \left\{ \begin{array}{lcl}
             \textrm{sgn}(x)G_{t}, &\textrm{if} &T_t\textless x\le T_{t+1}
            \\ 0,  &\textrm{if}  &|x|\le T_1
            \end{array}\right.,
    \end{equation}
    where $G_t\in {\mathbb{R}}^{+}$ is from a finite alphabet $\mathcal{G}=\{-G_l,\cdots,-G_1,0,G_1,\cdots,G_l\}$ which consists of $2l+1$ quantization levels and needs $\lceil{\log}_2(2l+1)\rceil$ bits. The elements in $\mathcal{G}$ satisfies $G_{t_1}\le G_{t_2}$ for any $t_1 \le t_2$. $T_t$ is a threshold from the set $\mathcal{T}=\{T_t\in{\mathbb{R}}^{+}:1\leq t \leq l+1\}$, and $T_t$ is defined as
    \begin{equation}
        T_t= \left\{ \begin{array}{lcl}
             \frac{1}{2}G_1, &\textrm{if} &t=1
            \\ \frac{1}{2}(G_{t-1}+G_t),  &\textrm{if} &2\leq t\leq l
            \\\infty, &\textrm{if} &t=l+1
            \end{array}\right..
    \end{equation}
    The thresholds of $Q_h(x)$ are usually set as
    \begin{equation}
    \begin{aligned}
    \mathcal{T}&=\{-T_{l+1},-T_{l},\cdots,-T_1,T_1,\cdots,T_{l},T_{l+1}\}\\
    &=\{-\infty,-lG+\frac{1}{2}G,\cdots,-\frac{1}{2}G,\frac{1}{2}G,\cdots,lG-\frac{1}{2}G,\infty\},
    \end{aligned}
    \end{equation}
     where $G$ is the step length between two adjacent thresholds, $lG-\frac{1}{2}G$ and $-lG+\frac{1}{2}G$ denote the upper and lower bounds, which are referred to as $G_{{b}}$ and $-G_{{b}}$ in the following for convenience.
    %\subsection{Soft Quantizer}
    \subsection{{QLcgNet}}
    \begin{figure}[t]
\vspace{-0.2cm}
\setlength{\belowcaptionskip}{-0.4cm}
\renewcommand{\captionfont}{\small}
\centering
\includegraphics[scale=.51]{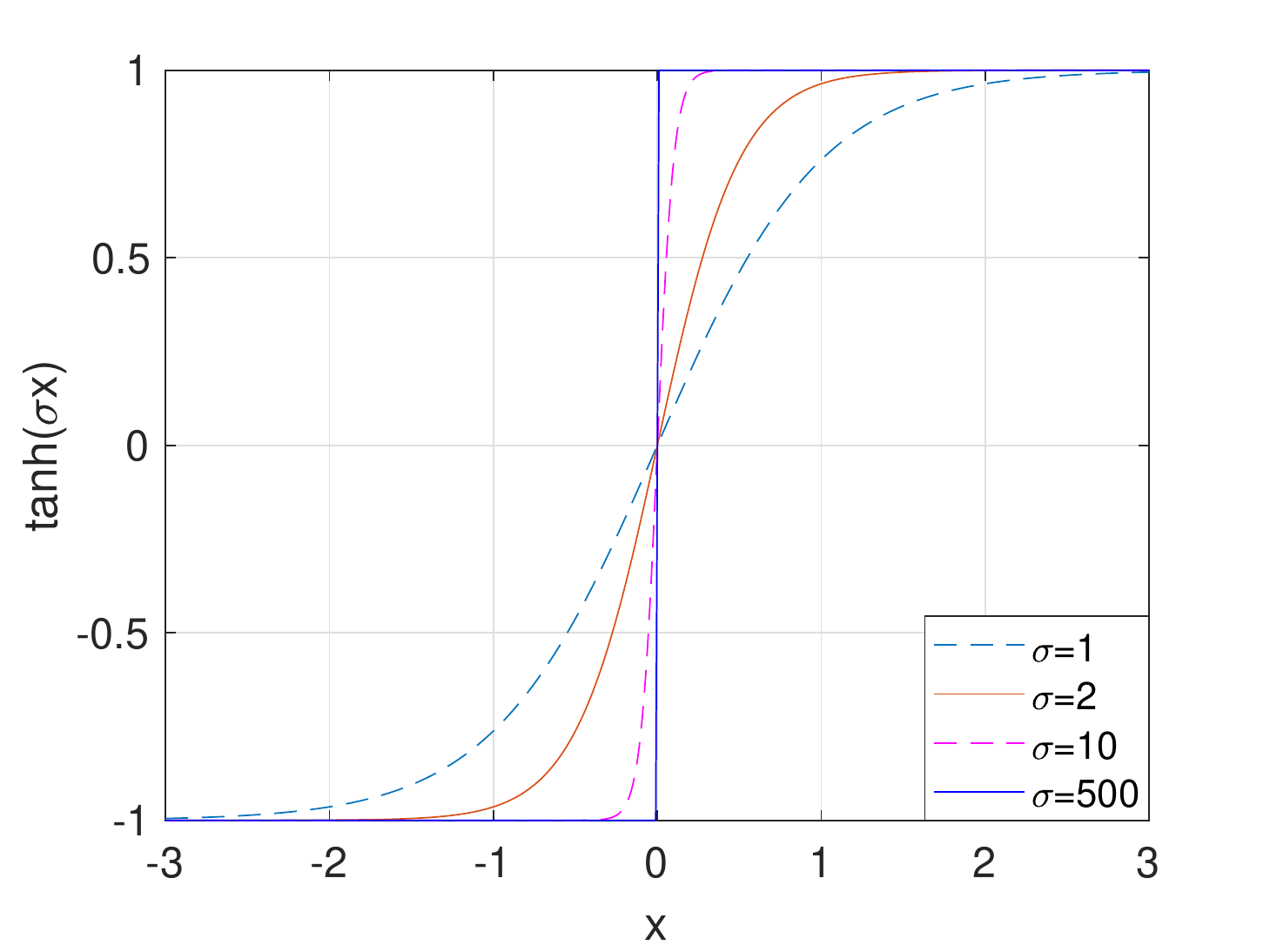}
\caption{The $\tanh$ function with different smoothing coefficients $\sigma$.}
\label{fig5}
\normalsize
\end{figure}
    Due to the various distributions of the network parameters, using a hard quantization function with uniformly distributed thresholds usually result in unexpected performance losses \cite{Wadayama2018Joint}. Therefore, in this subsection, we aim to design an adaptive soft staircase function which is able to minimize the MSE loss function with quantized parameters, i.e. $\mathcal{L}(\{Q({\bm{\Theta}}^{(1)}),\cdots,Q({\bm{\Theta}}^{(L)})\})$. {In other words, the stairs of a hard quantizer is set empirically by hand, while those of the proposed
soft quantizer can be learned by training with the aim of minimizing the NMSE.} Note that the hard staircase function (\ref{Eq25}) is not differentiable at the threshold points and its derivative is zero almost everywhere, this hinders the backpropagation process of the gradients. In order to overcome this difficulty and integrate an adaptive quantizer into LcgNetV directly, we propose a new soft staircase function $Q_s(\cdot)$ to approximate (\ref{Eq25}) with non-zero derivatives everywhere. The basic component of $Q_s(\cdot)$ is the following $\tanh(\cdot)$ function:
    \begin{equation}
    \tanh(x)=\frac{e^x-e^{-x}}{e^x+e^{-x}},
    \end{equation}
    which is a well-known activation function in the field of DL, and it is easy to obtain its derivatives.
    Fig. \ref{fig5} shows the curves of $\tanh(\sigma x)$ with different values of $\sigma$, where $\sigma$ can be interpreted as a smoothing coefficient. As can be seen, $\sigma$ controls the degree of smoothness of $\tanh(\sigma x)$,  and when $\sigma\rightarrow\infty$, $\tanh(\sigma x)$ can well approximate a hard staircase function with only one stair, i.e., the $\textrm{sgn}(\cdot)$ function.

\begin{figure}[h]
\vspace{-0.0cm}
\setlength{\belowcaptionskip}{-0.4cm}
\renewcommand{\captionfont}{\small}
\centering
\includegraphics[scale=.51]{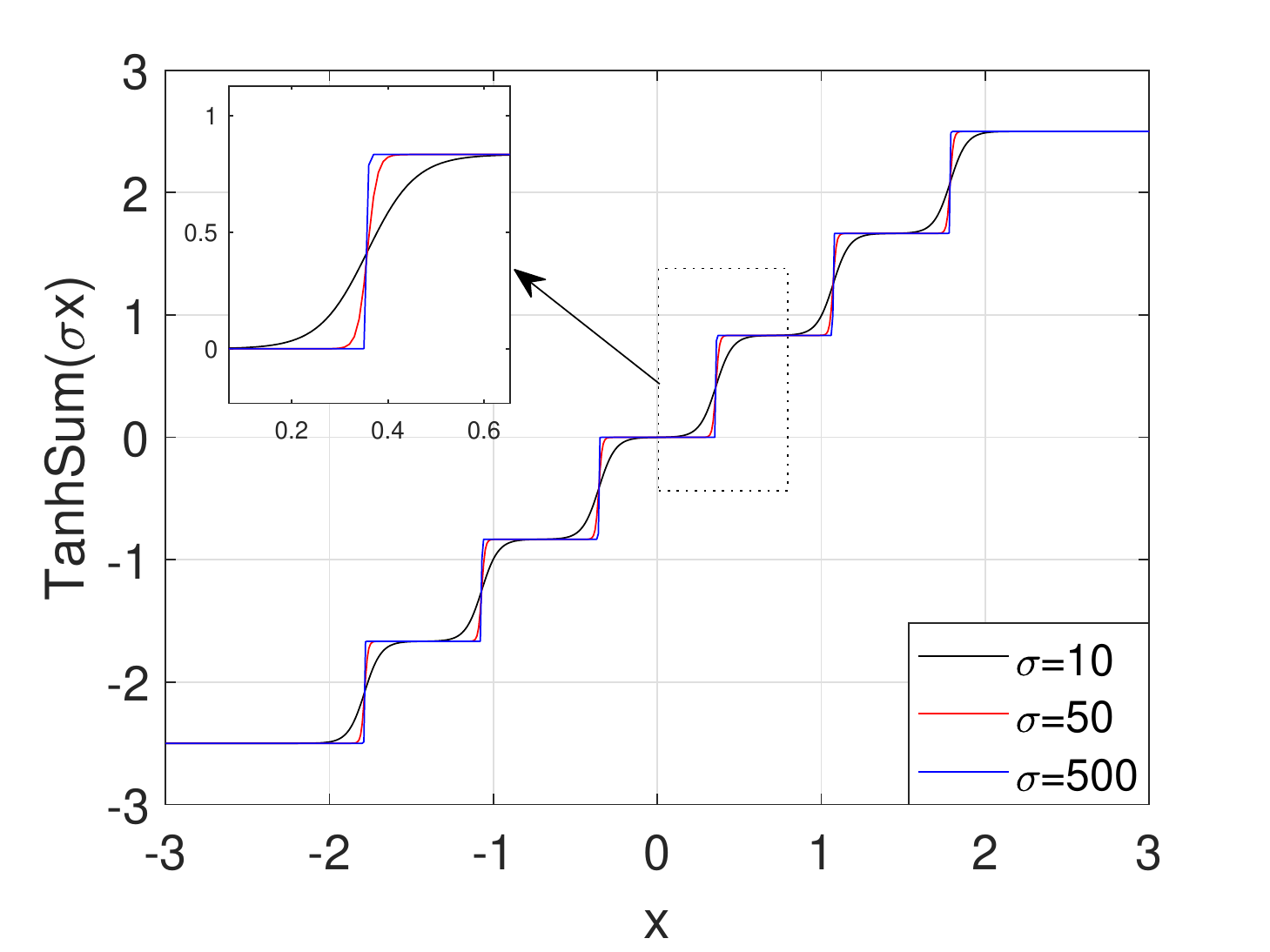}
\caption{The $3$-bit quantization TanhSum functions with different smoothing coefficients $\sigma$.}
\label{fig6}
\normalsize
\end{figure}

   Our soft staircase function, referred to as the \emph{TanhSum} function, consists of the summation of some $\tanh(\cdot)$ functions with different offsets. Its basic form is given by
    \begin{equation}
   \textrm{TanhSum}(x)=\frac{G_{{b}}}{2l}\sum_{t=1}^{2l}\tanh(\sigma(x+G_{{b}}-(t-1)G)),
    \end{equation}
which contains $2l+1$ stairs.
Fig. \ref{fig6} shows the curves of the proposed $\textrm{TanhSum}(\cdot)$ function with different values of $\sigma$ for 3-bit quantization. It can be observed that similar to the $\tanh(\cdot)$ function, the $\textrm{TanhSum}(\cdot)$ function gradually converges to a hard staircase function with the increasing of $\sigma$.
Furthermore, in order to endow the proposed $\textrm{TanhSum}(\cdot)$ function with the ability to learn the patterns and distributions of the network parameters, we introduce a set of learnable parameters $\bm{\Phi}=\{w_{1t},w_{2t},b_{1t},b_{2t}\}_1^{2l}$ to every component $\tanh$ function. As a result, we can obtain the following soft quantizer:
    \begin{equation}\label{Eq30}
      Q_s(x)=\sum_{t=1}^{2l}w_{1t}\tanh(\sigma(w_{2t}x+G_{{b}}-(t-1)G+b_{1t}))+b_{2t},
    \end{equation}
    where $w_{1t}$ and $w_{2t}$ are employed to adjust the length and height of the $t$-th level, $b_{1t}$ and $b_{2t}$ are the corresponding biases. %, which are initialized as $1,1,0,0$, respectively.
From (\ref{Eq30}), we can see that the entire input-output relationship of the proposed soft quantizer can be denoted by $Q_s(\cdot;\bm{\Phi},\{\sigma,l,G_{{b}}\})$, where $\{\sigma,l,G_{{b}}\}$ represent tunable hyperparameters which reflect the basic structure of the proposed soft quantizer.

\begin{figure}[t]
\vspace{0.2cm}
\setlength{\belowcaptionskip}{-0.4cm}
\renewcommand{\captionfont}{\small}
\centering
\includegraphics[scale=.26]{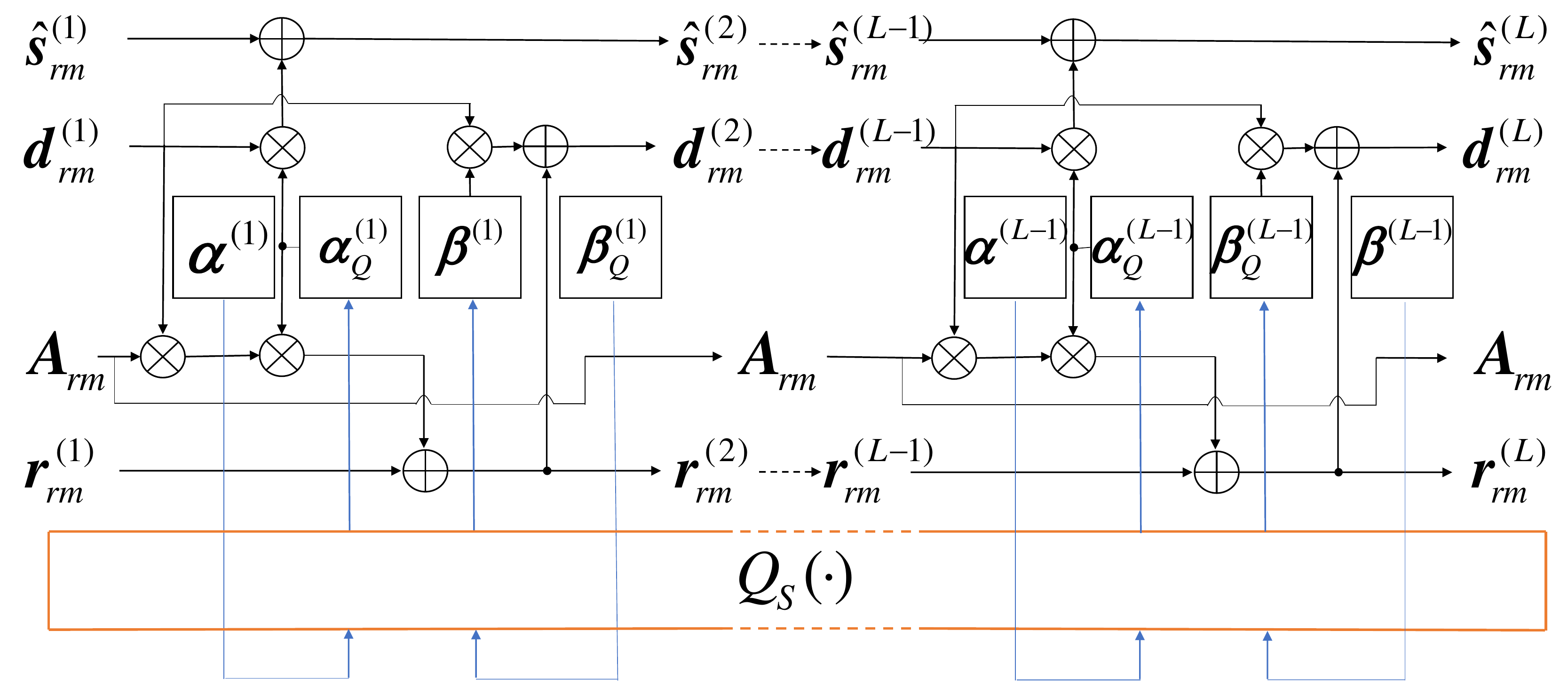}
\caption{The $i$-th layer of QLcgNetV with the proposed soft quantizer.}
\label{fig7}
\normalsize
\end{figure}

    The structure of the proposed QLcgNetV is shown in Fig. \ref{fig7}, where we integrate $Q_s(\cdot;\bm{\Phi},\{\sigma,l,G_{{b}}\})$ into the $L$-layer LcgNetV and use it to quantize the trained parameters ${\bm{\Theta}}_v$. It is important to note that there is only one quantizer $Q_s(\cdot)$ in the entire system, and it is used to quantize all the parameters. With given step-size parameters ${\bm{\Theta}}_v$, $\bm{\Phi}$ are optimized by minimizing the MSE loss ${\mathcal{L}}^{(L)}_{\textrm{Qs}}( \bm{\Phi}) $ from Eq. (\ref{j}) with the ``annealing strategy", i.e., we gradually increase the smoothing coefficient $\sigma$ to fintune $\bm{\Phi}$ and eventually the soft staircase function will converge to a discrete-valued staircase function.
\section{Numerical results}\label{Simulation}
In this section, we first evaluate the convergence property of the proposed networks. Then, the {detection} performance of LcgNetS and LcgNetV is presented under various channel models to demonstrate their advantages. {Next, we provide the performance
comparison between the proposed network and DetNet \cite{abs180507631} in terms of detection and complexity performances.} Finally, we provide the performance of QLcgNetV to show its advantages over that with a hard quantizer. In all our simulations, the definitions of NMSE and SNR are given by
\setcounter{TempEqCnt1}{\value{equation}}
\setcounter{equation}{25}
\begin{equation}
\mathrm{NMSE} = E\left\{ \frac{||{\hat {\bm{s}}} -\bm{ s }||^2}{{||\bm{s}||^2}}\right\},
\end{equation}
\begin{equation}\label{Eq32}
 \mathrm{SNR}=\frac{\mathbb{E}\{||\bm{s}||^2\}}{\mathbb{E}\{||\bm{n}||^2\}}.
\end{equation}

\subsection{Convergence Property}
\begin{figure}[t]
\vspace{0cm}
\setlength{\belowcaptionskip}{-0.4cm}
\renewcommand{\captionfont}{\small}
\centering
\includegraphics[scale=.51]{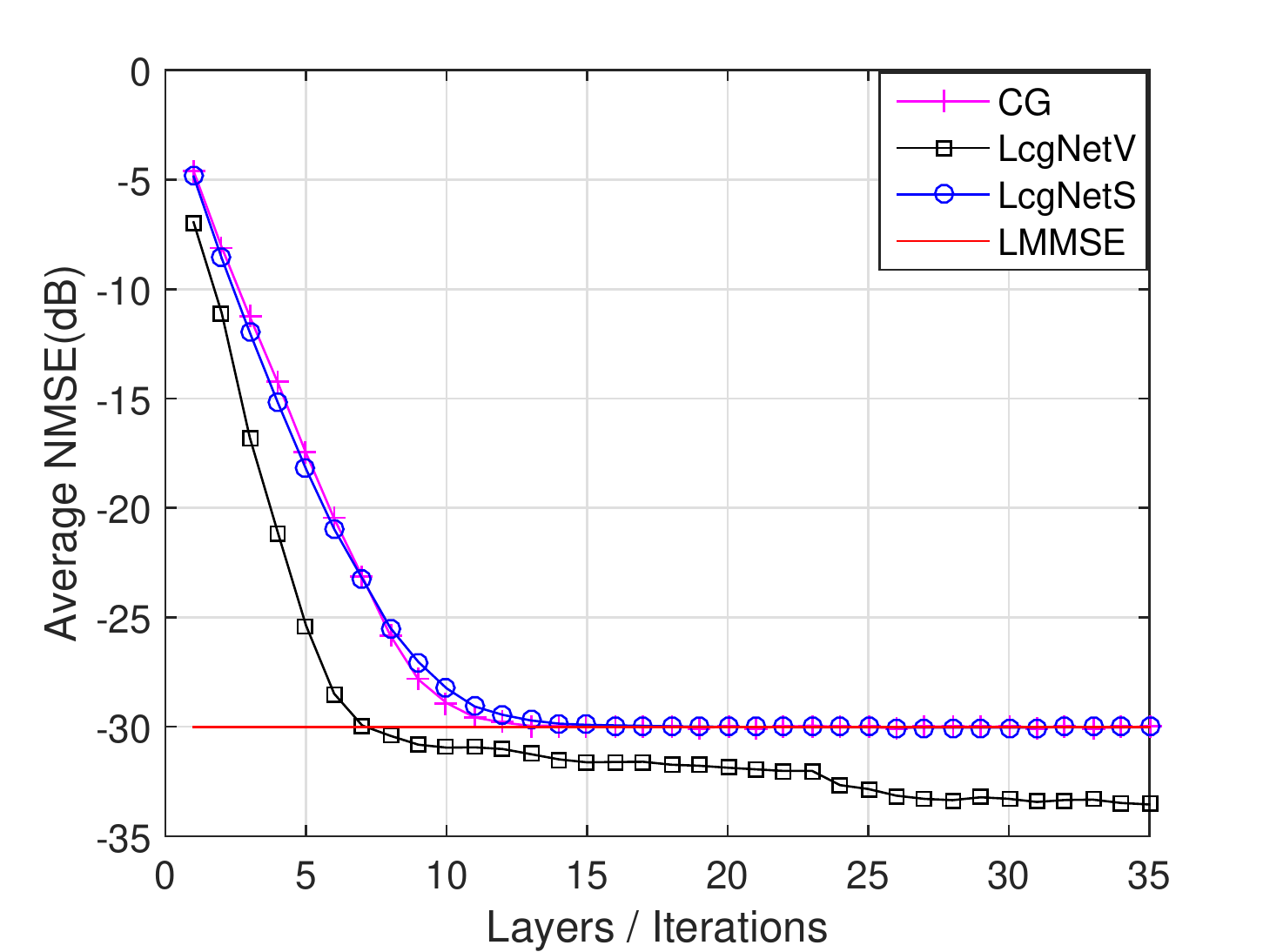}%CONV1
\caption{NMSE versus the number of layers or iterations when $\textrm{SNR}=30$ $\textrm{dB}$ in a $32\times 64$ MIMO system with Rayleigh fading channel model.}
\label{fig13}
\normalsize
\end{figure}
\begin{figure}[t]
\vspace{0cm}
\setlength{\belowcaptionskip}{-0.4cm}
\renewcommand{\captionfont}{\small}
\centering
\includegraphics[scale=.51]{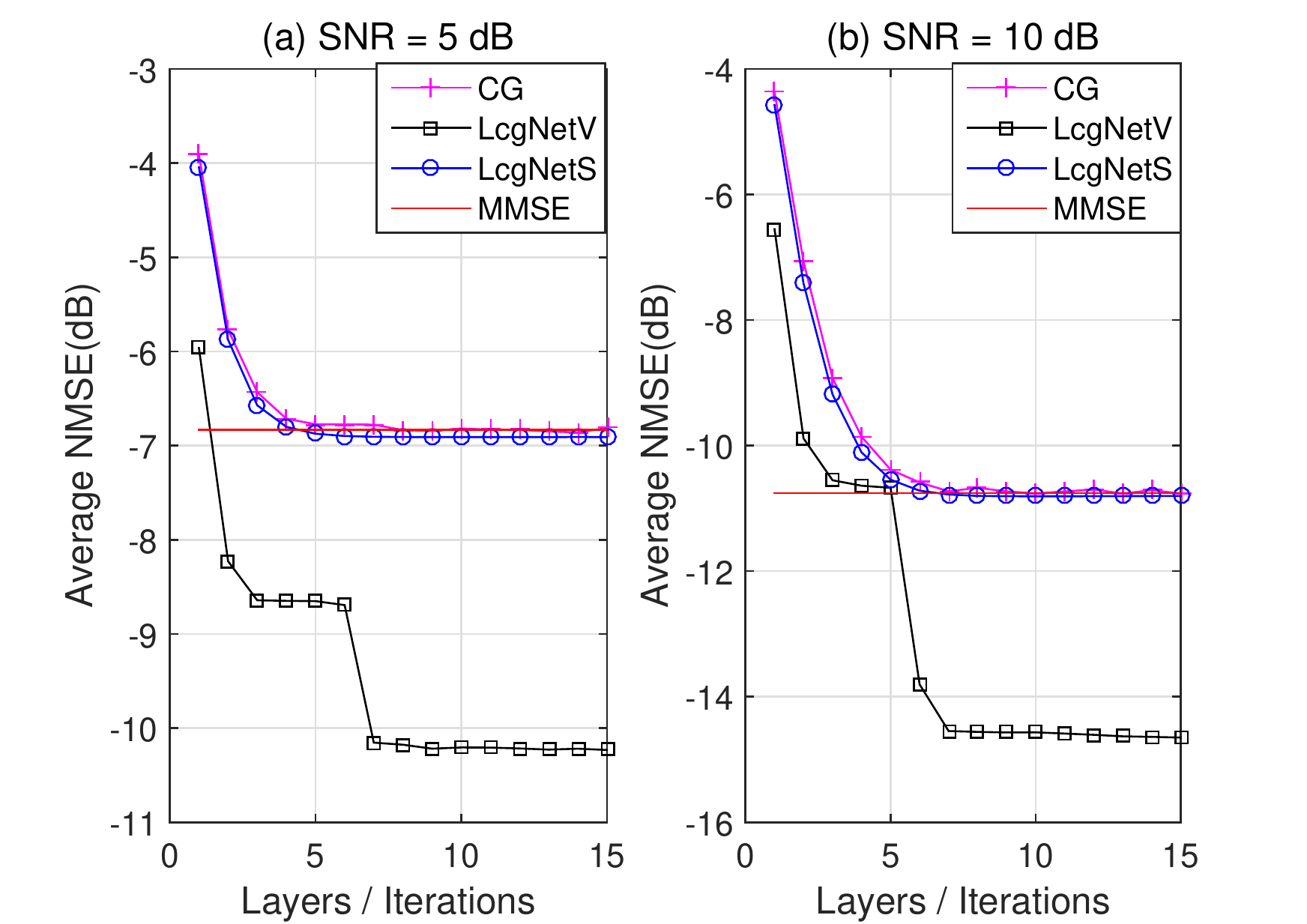}%CONV1
\caption{NMSE versus the number of layers or iterations when SNR=5 dB and 10 dB in a $32\times 64$ MIMO system with Rayleigh fading channel model.}
\label{fig3}
\normalsize
\end{figure}
 \newcounter{TempEqCnt2}
\setcounter{TempEqCnt2}{\value{figure}}
\setcounter{figure}{9}
\begin{figure*}[t]
\vspace{0.0cm}
\setlength{\belowcaptionskip}{-0.4cm}
\renewcommand{\captionfont}{\small}
\centering
\includegraphics[scale=.51]{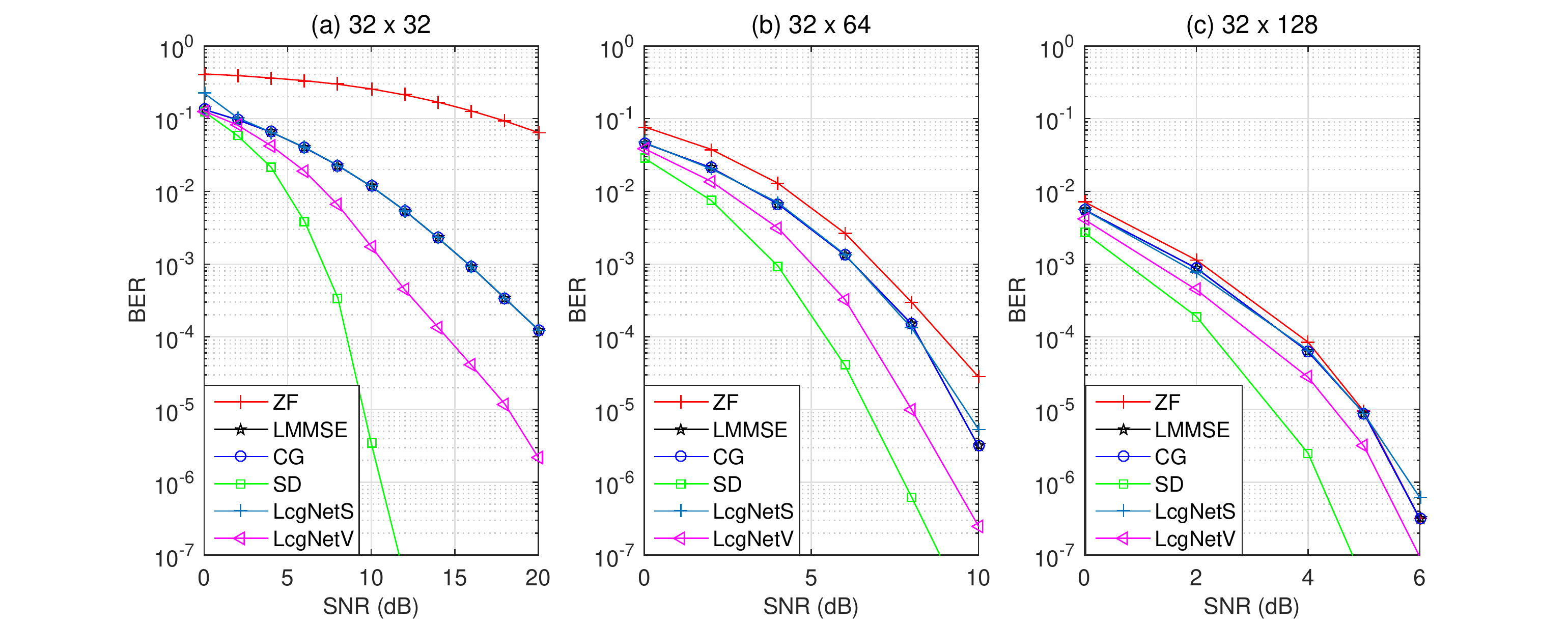}
\caption{{BER performance comparison between the considered detectors with Rayleigh fading channel model.}}
\label{fig4}
\normalsize
\end{figure*}

%\begin{figure*}[b]
%\vspace{0.0cm}
%\setlength{\belowcaptionskip}{-0.5cm}
%\renewcommand{\captionfont}{\small}
%\centering
%\includegraphics[scale=.40]{ber3.eps}
%\caption{BER performance comparison between the considered detectors.}
%\label{fig4}
%\normalsize
%\end{figure*}
In Fig. \ref{fig13} and Fig. \ref{fig3}, we investigate the performance of the proposed networks with different numbers of layers (or equivalently iterations), in terms of NMSE.
 We consider a $32\times 64$ MIMO system with Rayleigh fading channel,
where each element of the channel matrix $\bm{H}$ follows from a zero-mean Gaussian distribution with variance $1/N_t$ and {BPSK modulation is employed}.
It can be observed that the convergence speed of LcgNetS is similar to that of the CG detector, while LcgNetV with high dimensional step-sizes converges much faster than those two for various SNR levels. For example, LcgNetV takes 7 layers to achieve -30 $\textrm{dB}$ NMSE when SNR = 30 $\textrm{dB}$, while LcgNetS and the CG detector require 14 layers/iterations.
Furthermore, LcgNetV achieves a lower NMSE than LcgNetS and the CG detector with the same $L$, e.g., from Fig. \ref{fig13}, we can see that a 6.72 $\textrm{dB}$ NMSE gain can be achieved when $L=7$. Surprisingly, it can also be seen that the performance of LcgNetV with a few layers even exceeds that of the LMMSE detector (known as the performance limit of the CG detector). This performance gain over the LMMSE detector can partly be attributed to the power of offline supervised learning process, which endows the DL based method with more prior knowledge than the LMMSE detector, such as transmitted signals $\bm{s}$ (also the labels).
As shown in Fig. \ref{fig3}, when $L = 6$, LcgNetV achieves 3.35 $\textrm{dB}$ and 3.08 $\textrm{dB}$ performance gains over the LMMSE detector under SNR = 5 dB and 10 dB, respectively. It can also be observed from these two figures that in order to achieve the same performance with the LMMSE detector, more layers/iterations are needed with the increasing of SNR.\\

{
Furthermore, from Fig. \ref{fig3}, we can observe that for LcgNetV,  the NMSE decreases rapidly when the number of layers achieves a certain value.   This is a unique characteristic of the neural network. Note that a feedforward DNN can approximate any continuous function with arbitrary precision, providing that the numbers of layers and parameters are sufficiently large.
Therefore, the learning power of the proposed LcgNetV is constrained by the number of layers $L$. When $L$ gradually increases and reaches a large enough value, LcgNetV will be endowed with the ability to improve the detection performance, which further results to the {significant} decrease of the NMSE.   }

{It is worth noting that the required number of layers of the proposed networks is related  to the number of transmit antennas $N_t$, and we find by simulations that the detection performance would  gradually approach to saturation when the number of layers is larger than about $N_t/2$. }
% 30DB
%is used to measure the noise level, which is set as 30 $\textrm{dB}$ in this simulation. Each element of the channel matrix $\bm{H}$ obeys a zero-mean Gaussian distribution with variance $1/N_t$.
%Fig. \ref{fig3} compares the convergence property of the proposed networks with the CG-aided MMSE detection algorithm, which shows that the convergence property of the LcgNetS is same as the CG-aided MMSE and the LcgNetV with parameters of high dimensional step-size significantly outperforms all other algorithms. For instance, to reach the 30 $\textrm{dB}$ NMSE of the MMSE algorithm, the LcgNetV takes only $7$ layers yet the LcgNetS and the CG-aided MMSE take $14$ layers and iterations, respectively. Furthermore, with more layers, the LcgNetV exhibits better NMSE which is beyond the performance limit of the MMSE algorithm.
\subsection{{Detection} Performance}
\subsubsection{{Comparison with the ML detector}}
In this subsection, we first provide a BER performance comparison between LcgNetV, the ML detector and the LMMSE detector {under BPSK modulation} in Fig. \ref{figml}. Due to the limitation of computing power, we only consider a $10\times 10$  MIMO system with Rayleigh fading channel.
As can be seen, although the proposed LcgNetV outperforms the LMMSE detector greatly, there is still some gap to the optimal ML bound.
{This is mainly due to the fact that the proposed LcgNetV can be viewed as an improved version of  LcgNetS by introducing some certain level of nonlinearity, but its performance is still limited by the structure of the detection network for small scale MIMO systems. Intuitively, by improving the level of nonlinearity in LcgNetV, its performance can be further improved, which is left for future work.}
%This is mainly due to the fact that the proposed LcgNetV and the LMMSE detector are both linear detectors, and their performance is known to be limited for small scale MIMO systems. Therefore, by introducing a certain level of nonlinearity into LcgNetV, its performance may be further improved, which is left for future work.
\setcounter{TempEqCnt2}{\value{figure}}
\setcounter{figure}{8}
\begin{figure}[t]
\vspace{0cm}
\setlength{\belowcaptionskip}{-0.4cm}
\renewcommand{\captionfont}{\small}
\centering
\includegraphics[scale=.51]{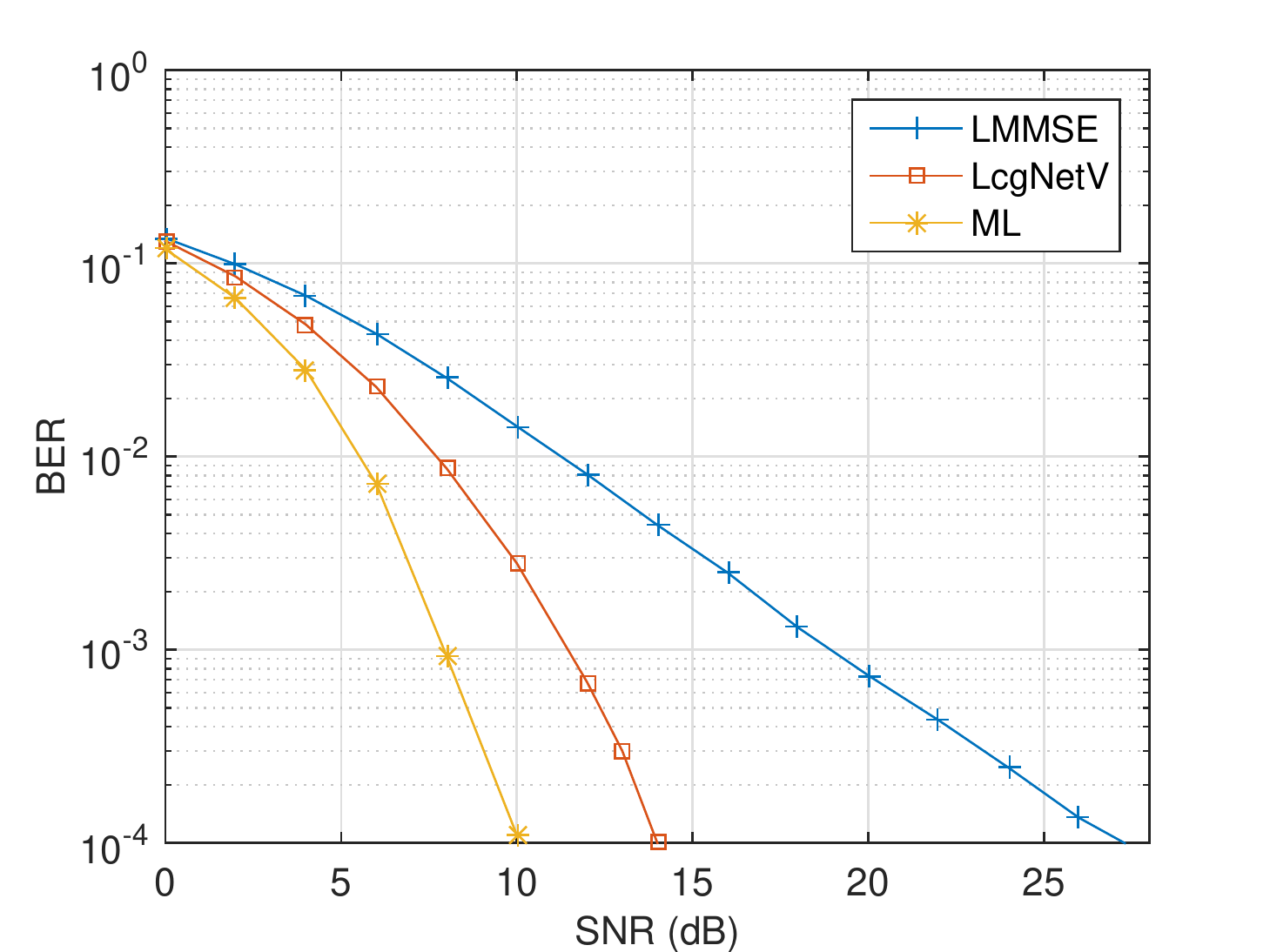}
\caption{{BER performance comparison between the considered detectors in a $10\times 10$ MIMO system with Rayleigh fading channel model.}}
\label{figml}
\normalsize
\end{figure}

\subsubsection{{Detection performance under different values of $N_r$}}
We investigate the BER performance of the proposed networks in large scale MIMO systems under Rayleigh fading channel,  where a clear channel-hardening phenomenon can be observed. Specifically,  in Fig. \ref{fig4}, we compare the BER performance of the proposed LcgNetS and LcgNetV with various existing MIMO detectors such as the ZF, LMMSE, CG and {SD detectors, and for the SD detector, the radius of the sphere is set to $2N_t\sigma^2$.} Here, three different system configurations are considered, i.e., $(N_t,N_r)=(32,32)$, $(32,64)$, $(32,128)$, {and BPSK modulation is employed}. The number of layers in the proposed networks is set to be $15$, and the iteration number of CG is fixed to $32$. {From Fig. \ref{fig4}, we can see that LcgNetS achieves the same performance as LMMSE, and LcgNetV outperforms the other detectors significantly in all three cases (except for the SD detector which can approach the performance of the ML detector).} Due to the advantages brought up by the channel-hardening phenomenon, the BER performance of all the considered detectors improves as the size of $\bm{H}$ increases. If we focus on the required SNR levels to reach BER=$10^{-4}$, the performance of LcgNetV improves about 0.38 $\textrm{dB}$ and 1.52 $\textrm{dB}$ in the $32\times 64$ and $32\times 128$ massive MIMO systems than that in the $32\times32$ system.
%Furthermore, the performance improvement achieved by linear detectors, such as ZF and LMMSE, is more significant than that by nonlinear detectors, e.g., SDR. For instance, in the $32\times 32$ system, the performance of ZF is inferior to that of the SDR detector, however in the $32\times 64$ system, ZF achieves a lower BER than SDR when $\textrm{SNR}\textgreater6$ $\textrm{dB}$, and in the $32\times 128$ system, ZF outperforms SDR for all the considered SNRs.
{\
\subsubsection{Detection performance under higher order modulations}

We further investigate the cases where higher order modulation schemes are employed, i.e., QPSK and 16QAM, and the results are shown in Fig. \ref{fig15}. A $32\times 64$ MIMO system is considered under Rayleigh fading channel and the number of layers of the proposed networks are fixed to $15$.
 It can be observed that {when QPSK is employed, LcgNetV can still outperform the baseline LMMSE detector in terms of the symbol error rate (SER) performance, but the performance gain is smaller than that under  BPSK modulation. When 16QAM is used, the SER performance of all the considered detectors (except for the SD detector) becomes similar. This is due to the fact that using higher order modulations  increases the difficulty of data detection, and the simple structure of LcgNet limits its ability to achieve better performance. However, it is worth noting that although the performance gain of LcgNetV over the LMMSE detector vanishes with the increasing of modulation order, its computational complexity is still lower.  In order to improve the performance of the proposed networks under high order modulation schemes, some additional modifications are required to be introduced into LcgNet, such as the utilization of multi-level activation functions as in \cite{abs-1812-01571}, and this is left for future work.}
% It can be observed that when QPSK is employed, LcgNetV can still outperform the baseline LMMSE detector in terms of the symbol error rate (SER) performance, but the detection performance gain is not as large as the cases with BPSK modulation. Under 16QAM, the SER performance of all the considered detectors except for the SD detector tends to be similar. It is worth noting that although the performance gain of LcgNetV over the LMMSE detector vanishes with the increasing of modulation order, its computational complexity is lower than the LMMSE detector.  In order to apply the proposed networks to high order modulation with better detection performance, some additional modifications are required to be introduced into LcgNet, such as the utilization of multi-level activation functions as \cite{abs-1812-01571}, and this is left for future work.

}
%\setcounter{TempEqCnt2}{\value{figure}}
%\setcounter{figure}{10}
%\begin{figure}[t]
%\vspace{0cm}
%\setlength{\belowcaptionskip}{-0.5cm}
%\renewcommand{\captionfont}{\small}
%\centering
%\includegraphics[scale=.45]{QPSKper1.eps}
%\caption{SER performance comparison of LcgNetV, LcgNetS and the LMMSE detector in $32\times 64$ MIMO and $32\times 128$ MIMO systems under QPSK modulation. }
%\label{fig15}
%\normalsize
%\end{figure}
\setcounter{TempEqCnt2}{\value{figure}}
\setcounter{figure}{10}
\begin{figure}[t]
\vspace{0cm}
\setlength{\belowcaptionskip}{-0.4cm}
\renewcommand{\captionfont}{\small}
\centering
\includegraphics[scale=.51]{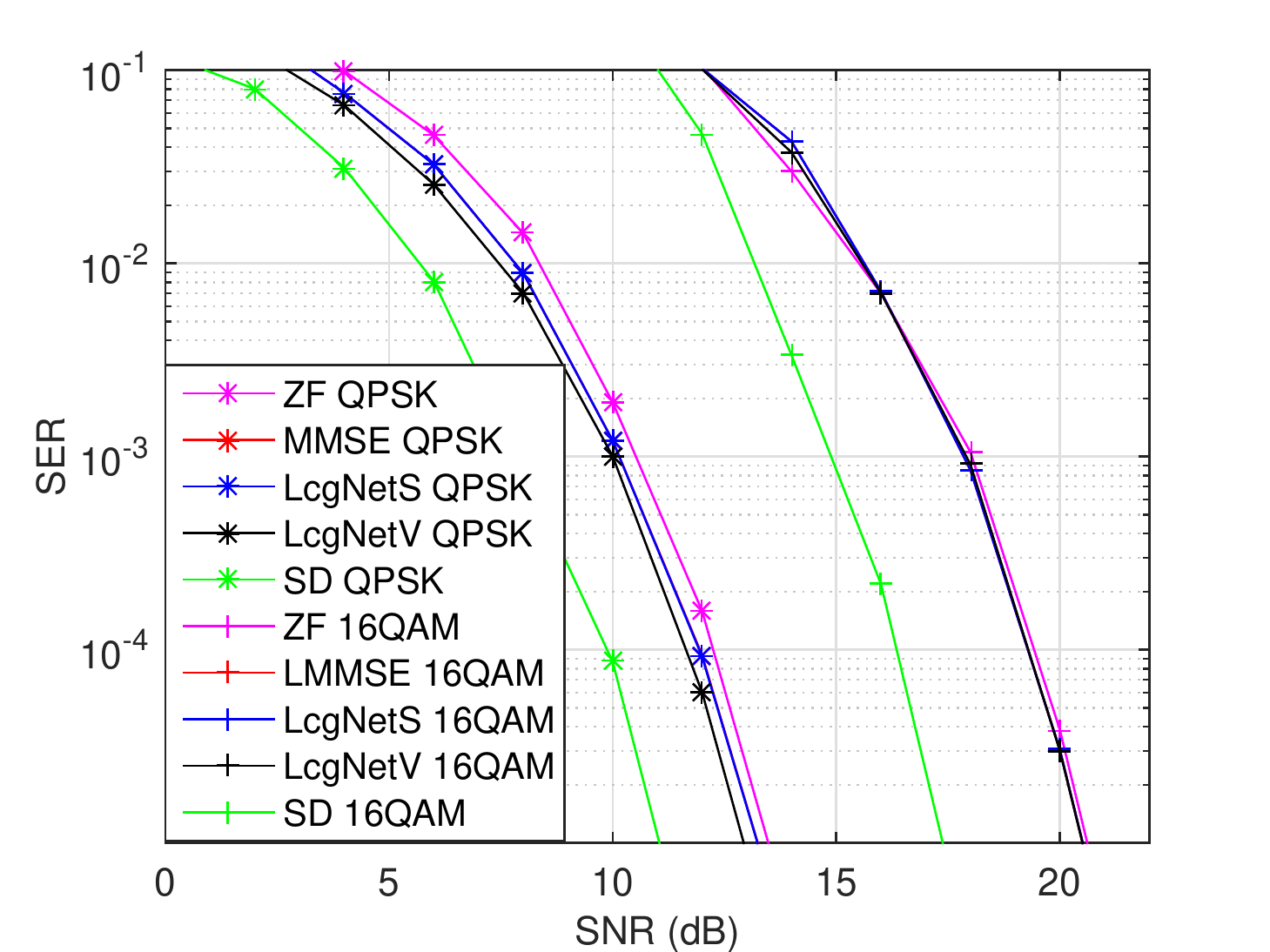}
\caption{{SER performance comparison between the considered detectors in a $32\times 64$ MIMO system under QPSK and 16QAM modulations.}}
\label{fig15}
\normalsize
\end{figure}

%Next, we investigate the BER performance in QPSK modulation and compare with the DetNet \cite{} of the proposed network
{
\subsubsection{Detection performance under correlated channel models}
We investigate the BER performance of the proposed detectors in some more realistic channel models, e.g., a spatial correlated channel model, which is known to be difficult for signal detection \cite{951380}, and a TDL-A MIMO  channel model  \cite{3GPP38}.}  The numbers of antennas at the transmitter and receiver are set to $(N_t, N_r) = (32,64)$ {and BPSK modulation is used}.
For the spatial correlated channel model, the channel matrix can be written in the following kronecker product form:
  \begin{equation}\label{spa}
  \bm{H}={\bm{R}}_r^{1/2}\bm{U}{\bm{R}}_t^{1/2},
  \end{equation}
  where $\bm{U}\in {\mathbb{C}}^{N_r\times N_t}$ obeys Rayleigh distribution, ${\bm{R}}_r \in {\mathbb{C}}^{N_r\times N_r}$ and ${\bm{R}}_t\in {\mathbb{C}}^{N_t\times N_t}$ denote the correlation matrices at the receiver and transmitter, respectively. ${\bm{R}}_t$ and ${\bm{R}}_r$ are generated from the exponential correlation model \cite{951380},
    whose component $r_{ij}$ can be written as
    \begin{equation}\label{exp}
             r_{ij}= \left\{ \begin{array}{lcl}
             r^{j-i}, &\textrm{if} &i\le j
            \\ r_{ji}^*,  &\textrm{if}  &i\textgreater j
            \end{array}\right.,
    \end{equation}
    where $|r|\leq 1$ is the correlation coefficient of neighboring receive branches and it is set to be 0.5 in this work.
   {In addition, the TDL-A MIMO  channel model is a non-line-of-sight (NLOS) link level channel model form the 3GPP TR 38.901 standard \cite{3GPP38}. The channel matrices of the TDL-A  MIMO channel model are generated  using the Matlab (2019a) 5G Toolbox, where the transmit and receive correlation matrices are set according to  the exponential correlation model in   \eqref{exp} and the correlation coefficient is set to 0.5. Other channel parameters are listed in Table \ref{Tab2}.}
   {
   \begin{table}[h]
\vspace{-0.0em}
\setlength{\abovecaptionskip}{-0.1cm}
\setlength{\belowcaptionskip}{-0.1cm}
\caption{Channel Model Configuration}
\begin{center}
\begin{tabular}{c|c}
\hline
\textbf{Parameter}& \textbf{Value}\\
\hline
Fading process statistical distribution & Rayleigh \\
\hline
Antenna polarization arrangement & Co-Polar\\
\hline
Desired RMS delay &  $3e^{-8}$s \\
\hline
 Maximum doppler shift & 5Hz\\
\hline
Sample rate of input signal & 30.72MHz \\
\hline
 Cross-polarization power ratio & 10dB \\
 \hline
 Desired K-factor for scaling & 9dB\\
 \hline
 Number of modeling sinusoids & 48\\
\hline
Normalize path gains & True \\
\hline
\end{tabular}
\label{Tab2}
\end{center}
\end{table}
    }
   %  Fig. \ref{fig14} shows the amplitudes of ${\bm{H}}^H\bm{H}$ in the correlated and Rayleigh channel models and it can be seen that the spatial correlation property undermines the channel-hardening phenomenon introduced by large scale antennas. Compared with the Rayleigh channel model, the matrix ${\bm{H}}^H\bm{H}$ in correlated channel model has smaller diagonal elements and larger off-diagonal ones, and its average amplitude difference between diagonal and off-diagonal elements is much smaller.

%    \begin{figure}[t]
%\vspace{0cm}
%\setlength{\belowcaptionskip}{-0.2cm}
%\renewcommand{\captionfont}{\small}
%\centering
%\includegraphics[scale=.35]{CR_comp.eps}
%\caption{An illustration of the differences between the exponential correlated and Rayleigh fading channel model.}
%\label{fig14}
%\normalsize
%\end{figure}
\begin{figure}[t]
\vspace{0cm}
\setlength{\belowcaptionskip}{0.0cm}
\renewcommand{\captionfont}{\small}
\centering
\includegraphics[scale=.51]{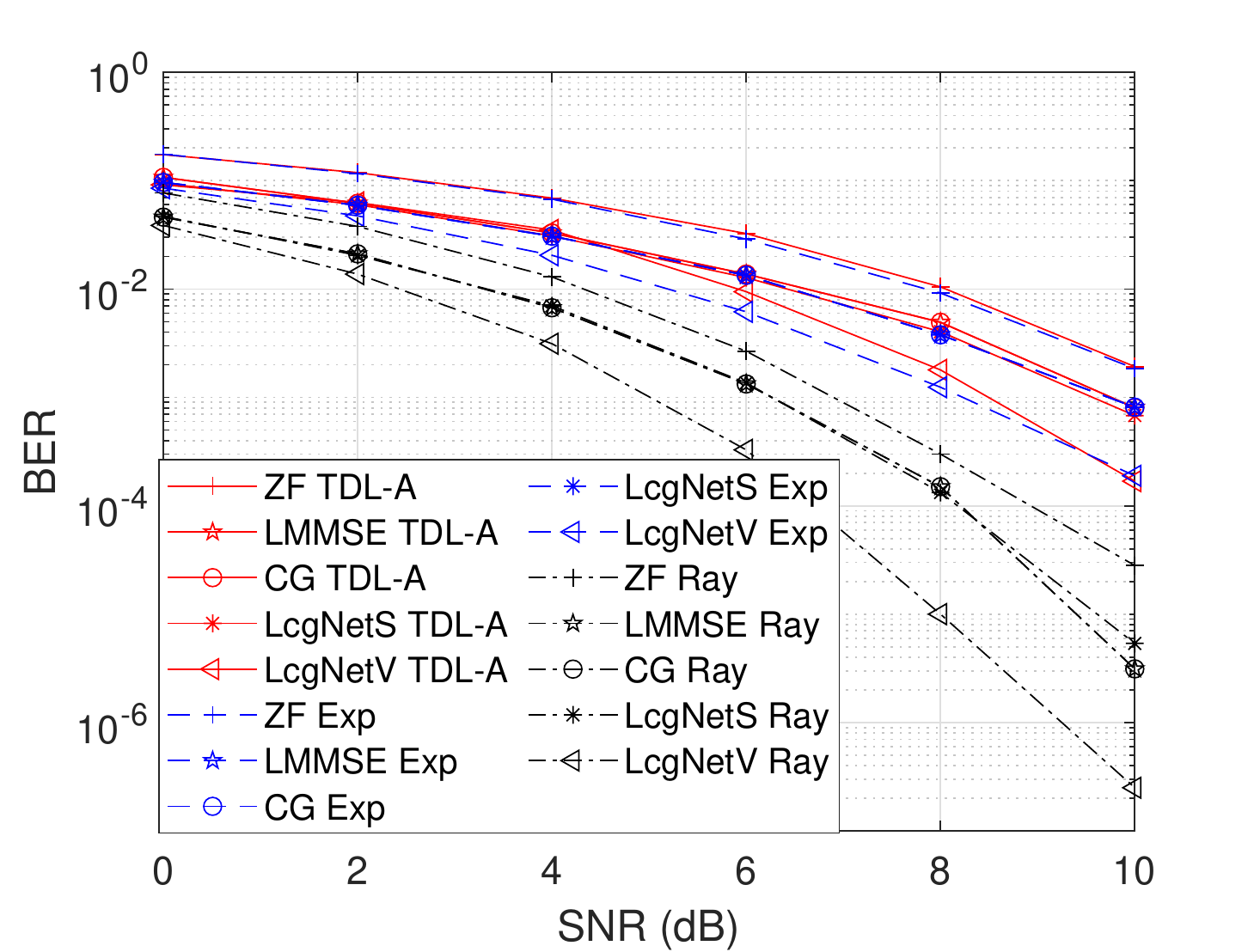}
\caption{{BER performance comparison between the considered detectors in a $32\times 64$ MIMO system with different channel models, where \emph{TDL-A},  \emph{Exp} and \emph{Ray} denote the  TDL-A MIMO channel, the spatial correlated channel and the Rayleigh fading channel, respectively.}}
\label{fig11}
\normalsize
\end{figure}

\begin{figure}[t]
\vspace{-0.10cm}
\setlength{\belowcaptionskip}{-0.4cm}
\renewcommand{\captionfont}{\small}
\centering
\includegraphics[scale=.51]{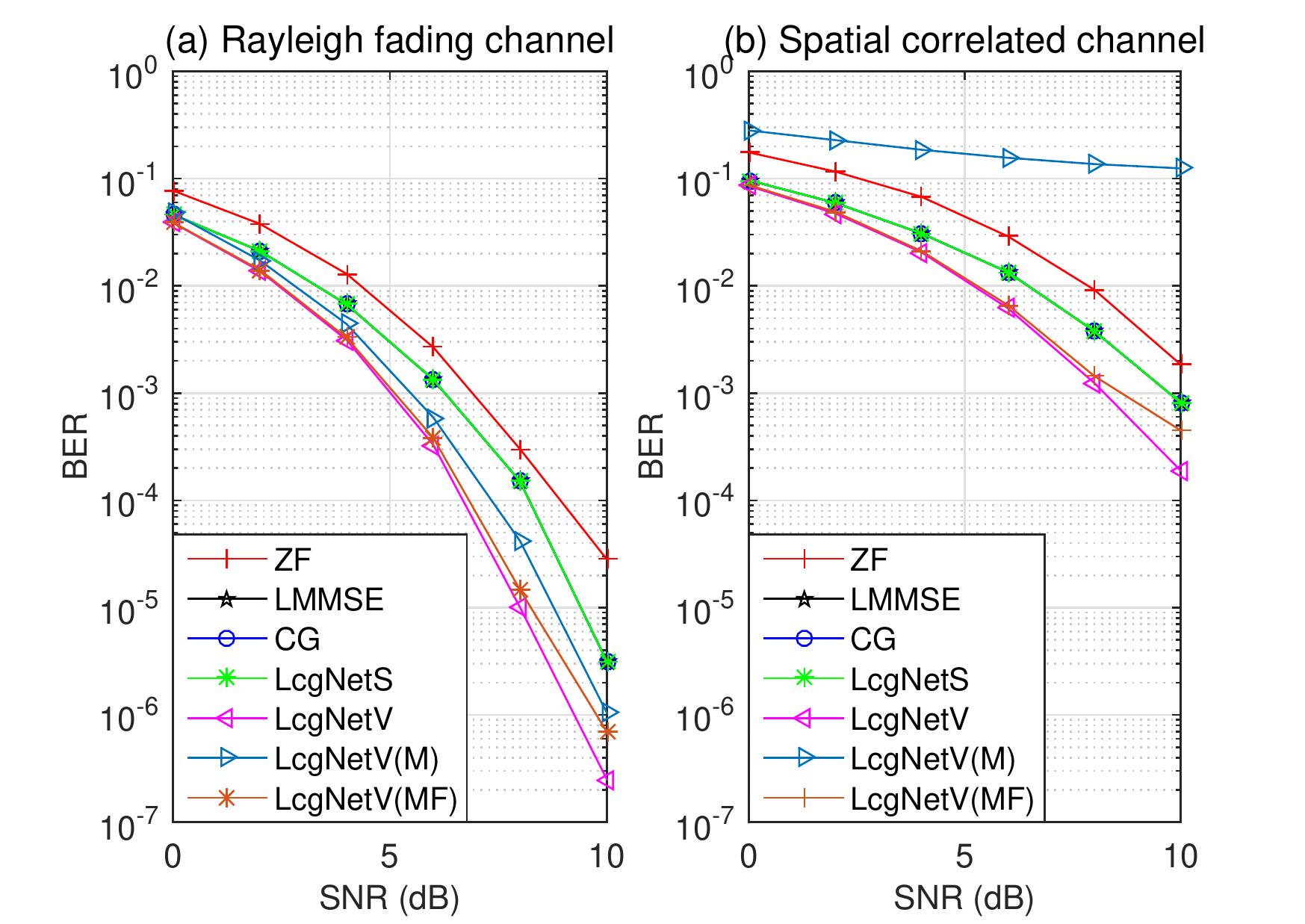}
\caption{{BER performance comparison between the considered detectors when the training and testing channel models are different.}}
\label{mismathch}
\normalsize
\end{figure}

 {In Fig. \ref{fig11}, we present the BER performance of the considered detectors in the aforementioned correlated channel models. We can observe that all the considered detectors experience a certain degree of performance degradation under the spatial correlated channel model and the TDL-A MIMO channel model. With the same  correlation matrices, LcgNetV exhibits almost the same  performance under the TDL-A MIMO channel model and the spatial correlation channel model \eqref{spa}. In these channel models,  LcgNetV achieves the best performance among all the considered counterparts.

Furthermore, in order to investigate the robustness of LcgNetV, we plot the BER performance of LcgNetV when the training  and  testing channel models are different in Fig. \ref{mismathch}. The resulting networks are referred to as LcgNetV (M). Specifically, LcgNetV (M) in Fig. \ref{mismathch} (a) and Fig. \ref{mismathch} (b) represent the networks that are trained under the spatial correlated and Rayleigh fading channel models, respectively.
It can be observed that model mismatch results to certain performance losses. In Fig. \ref{mismathch} (a), LcgNetV (M) still exhibits  superior performance as compared with the LMMSE detector under the Rayleigh fading channel model, while in Fig. \ref{mismathch} (b), LcgNetV (M) suffers from severe performance loss when tested under the spatial correlated channel model.  This suggests that when the statistical information of the channel is time-varying, the proposed networks should be finetuned periodically to avoid the potential performance loss. To verify this, we finetune LcgNetV (M)  under the testing channel models and the resulting networks are name as LcgNetV (MF) in Fig. \ref{mismathch}. It can be seen that although LcgNetV (MF) is trained under model mismatch, further finetuning it under the testing channel model can effectively compensate the performance loss.}

% Furthermore, in order to investigate the robustness of the LcgNetV, Fig. \ref{mismathch}  shows the BER performance of the LcgNetV when model mismatch happened, and the resulting networks are referred to as LcgNetV (M). Specifically, the LcgNetV (M) in Fig. \ref{mismathch} (a) and Fig. \ref{mismathch} (b) represent the networks which are trained under the correlated and Rayleigh channel model, respectively.
%It can be observed that the model mismatch could result to performance loss in both Rayleigh and correlated channel models. In Fig. \ref{mismathch} (a), LcgNetV (M) still exhibits  superior performance compared with the LMMSE detector under the Rayleigh channel model, while in Fig. \ref{mismathch} (b), LcgNetV (M) suffers from severe performance loss when tested under the correlated channel model.  To sum up, not only the difference between the statistic characteristic of the training and testing channel models but also the correlation characteristic of the testing channel model affect the performance loss caused by model mismatch. However, funtuning the proposed network under the testing channel model can compromise the performance loss effectively.

%Further, it is interesting to note that the proposed networks have fewer layers than DetNet (30 layers) \cite{abs180507631}, leading to much improved test time computational efficiency. This can be attributed to the fact that, the CG algorithm that we try to unfold is known for its fast convergence, while the DetNet tries to unfold the iteration of the PG algorithm, which is much slower (in terms of number of iterations required to achieve optimal solutions).
{
\subsection{{Performance Comparison with DetNet}}
In this subsection, we provide a detailed  performance/complexity/storage comparison between the proposed LcgNetV and the DetNet in \cite{abs180507631}.

First, the SER performance is compared  under QPSK modulation and Rayleigh fading channel, as show in Fig. \ref{det}. In LcgNetV, the number of layers is set to 15.
In DetNet, the sizes of ${\bm{z}}_k$ and ${\bm{v}}_k$ are set to $4N_t$ and $2N_t$, and the number of layers is set to 30 according to \cite{abs180507631}. In the simulation, the training process of LcgNetV and DetNet took about 2 hours and 3 days, respectively. It can be observed that the proposed network can achieve better performance when the scale of the MIMO system is large (e.g. in the $32\times128$ MIMO system), however its performance is inferior to that of  DetNet when $N_r$ is small. {This is because the proposed network unfolds the CG detector to improve the performance of the LMMSE detector, which is near-optimal when $N_r$ is very large.} For small-scale MIMO systems, the performance of the CG detector is not competitive and although LcgNetV can outperform the CG detector, its detection performance is still limited by the fixed structure of the CG iterations.
 \begin{figure}[t]
 \vspace{-0.0em}
 \setlength{\belowcaptionskip}{-0.4cm}
 \renewcommand{\captionfont}{\small}
  \centering
  \includegraphics[scale=.51]{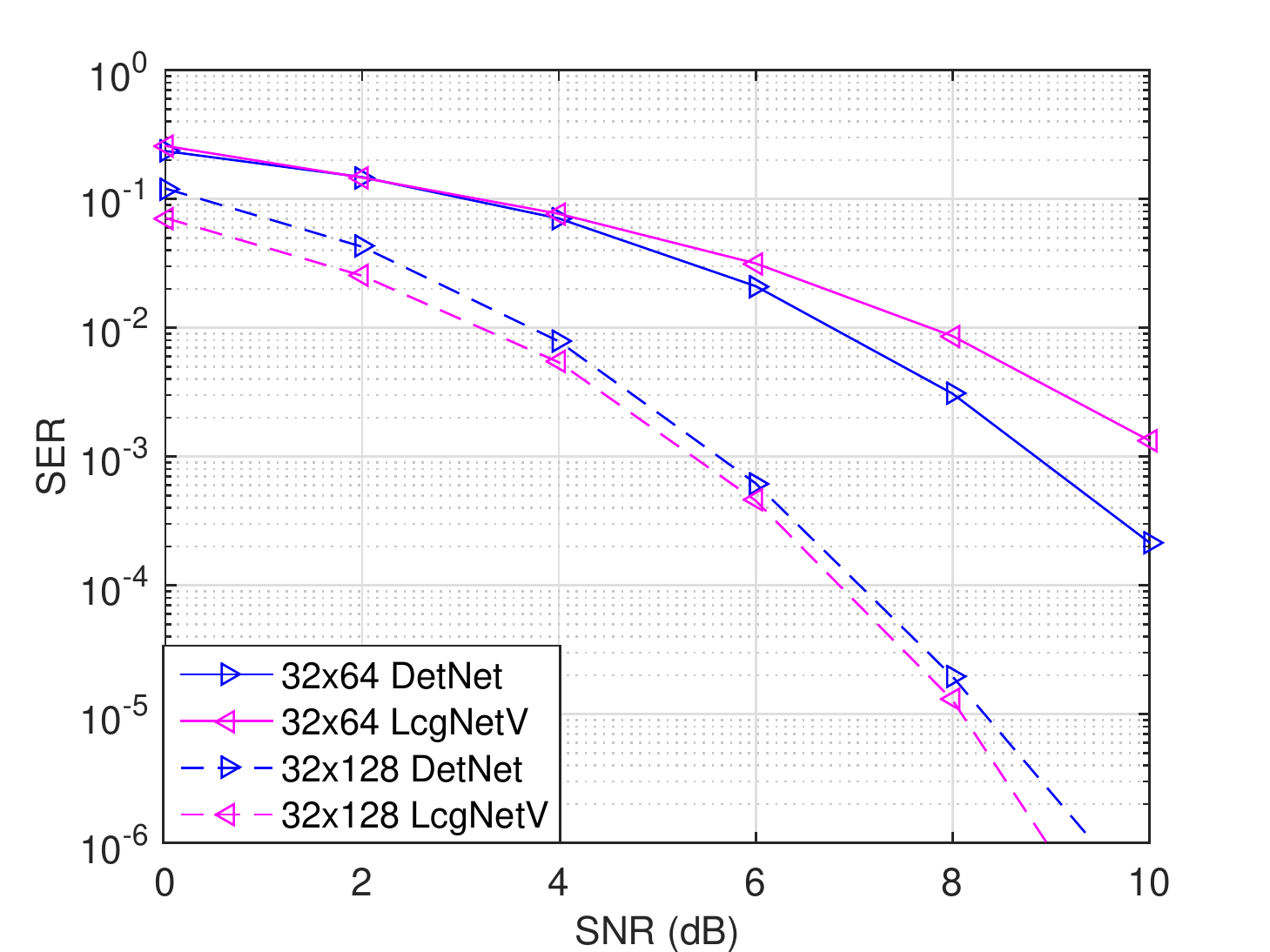}
  \caption{{SER performance comparison between DetNet and LcgNetV under QPSK modulation and Rayleigh fading channel.}}
  \label{det}
\end{figure}

Second, we investigate the computational complexities of LcgNetV and DetNet. Let $L_{\textrm{DetNet}}$ and $L_{\textrm{LcgNetV}}$ denote the numbers of layers required by DetNet and LcgNetV, respectively. According to [24], the computational complexity of DetNet is $\mathcal{O}((36N_t^2+4N_t)L_{\textrm{DetNet}})$, where we have ignored the nonlinear operations for simplicity. For LcgNetV, its complexity is on the order of $\mathcal{O}((4N_t^2+6N_t)L_{\textrm{LcgNetV}})$.
%More specifically, the numbers of real multiplication involved in the detection process of the DetNet and LcgNetV which
 Therefore, we can observe that although the complexity caused  by the nonlinear functions in DetNet are not taken into consideration, its complexity is still much higher than that of the proposed networks.

Finally, we compare the memory costs of LcgNetV and DetNet based on the number of bits required for
storing the network parameters, where we assume that $B$ bits are required to store one real number. Consider the $i$-th layer, the parameters $\left\{\bm{W}_{1 i}, \bm{b}_{1 i}, \bm{W}_{2 i}, \bm{b}_{2 i}, \bm{W}_{3 i}, \bm{b}_{1 i}, \delta_{1 i}, \delta_{2 i}\right\}$ and $\left\{{\bm{\alpha}}_r^{(i)},{\bm{\beta}}_r^{(i)}\right\}$ are required to be stored in DetNet and LcgNetV, respectively. Therefore, their  memory costs can be expressed as $(32N_t^2+8N_t+2)L_{\textrm{DetNet}}B$ bits and $4N_tL_{\textrm{LcgNetV}}B$ bits, respectively.

To summarize, we can see that the proposed LcgNetV is more suitable for massive MIMO systems due to its better SER performance, lower computational complexity and less memory cost.
}
\subsection{The Performance of QLcgNetV}
We then investigate the effects of the proposed soft quantizer on LcgNetV, i.e., the performance of QLcgNetV, in a $32\times 64$ Rayleigh fading channel {when BPSK is employed}.
The BER performance of QLcgNetV with both hard and soft quantizers are investigated, and the number of quantization bits are chosen as 3 and 4.
Given the prior knowledge of $G_{{b}}$ obtained by LcgNetV and the required number of quantization levels, we complete the training process of QLcgNetV in three steps and the smoothing coefficient $\sigma$ used in each step is from the set $\{10,50,100\}$ in an increasing manner. In each training step, the learning rate is chosen from $\{1e^{-4},5e^{-5},1e^{-5}\}$ successively and the training is terminated when the average NMSE stops decreasing. The adjustable parameters $\{w_{1t},w_{2t},b_{1t},b_{2t}\}$ in (\ref{Eq30}) are initialized to $\{1,1,0,0\}$ and the training data set contains $10^4$ samples.

\begin{figure}[t]
\vspace{0.0cm}
\setlength{\belowcaptionskip}{-0.4cm}
\renewcommand{\captionfont}{\small}
\centering
\includegraphics[scale=.51]{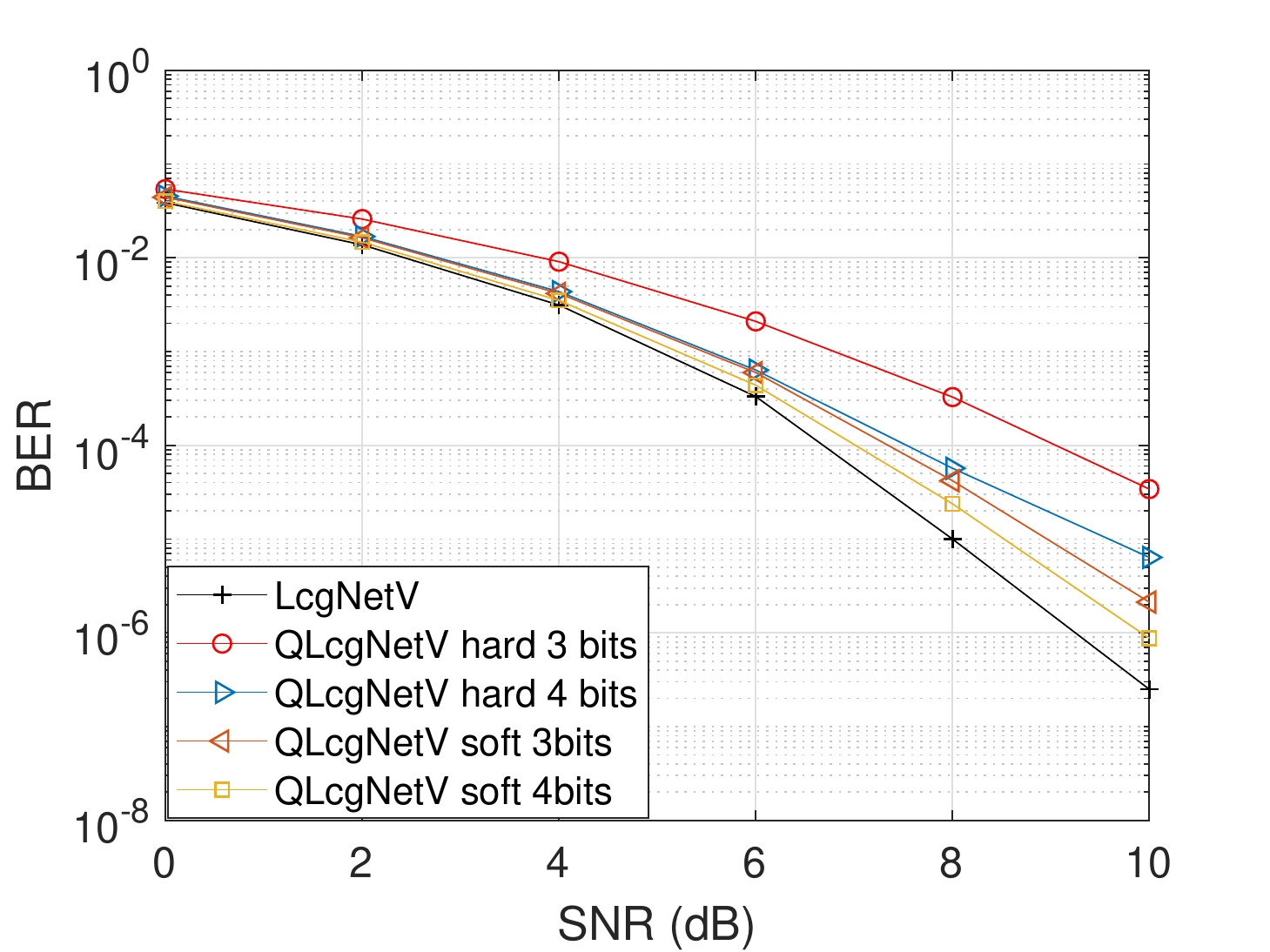}
\caption{BER performance comparison between LcgNetV and QLcgNetV with hard and soft quantizers ($32\times 64$, Rayleigh fading channel).}
\label{fig8}
\normalsize
\end{figure}

As shown in Fig. \ref{fig8},  QLcgNetV  significantly outperforms LcgNetV with a conventional hard quantizer and the BER performance achieved by a 3-bit soft quantizer is even lower than that achieved by a 4-bit hard quantizer.
The performance of QLcgNetV with 3-bit and 4-bit soft quantizers is only 0.55 dB and 0.27 dB away from that without quantization at BER=$10^{-4}$. This indicates that the network parameters in LcgNetV can be effectively compressed with minor performance loss. Specifically, if we assume that 32 bits are needed to store one real number without quantization and the number of layers is 15, then the memory cost of LcgNetV for storing the network parameters can be reduced from 61440 bits to 5760 bits if we employ QLcgNetV with a 3-bit soft quantizer. Also, it can be seen that there is a tradeoff between BER performance and memory cost, which mainly depends on system performance requirements and implementation.
%This illustrates that the new proposed quantization method enables the LcgNetV to achieve better BER performance even with less memory cost.

\begin{figure}[t]
%\vspace{-0.5cm}
\setlength{\belowcaptionskip}{-0.4cm}
\renewcommand{\captionfont}{\small}
\centering
\includegraphics[scale=.51]{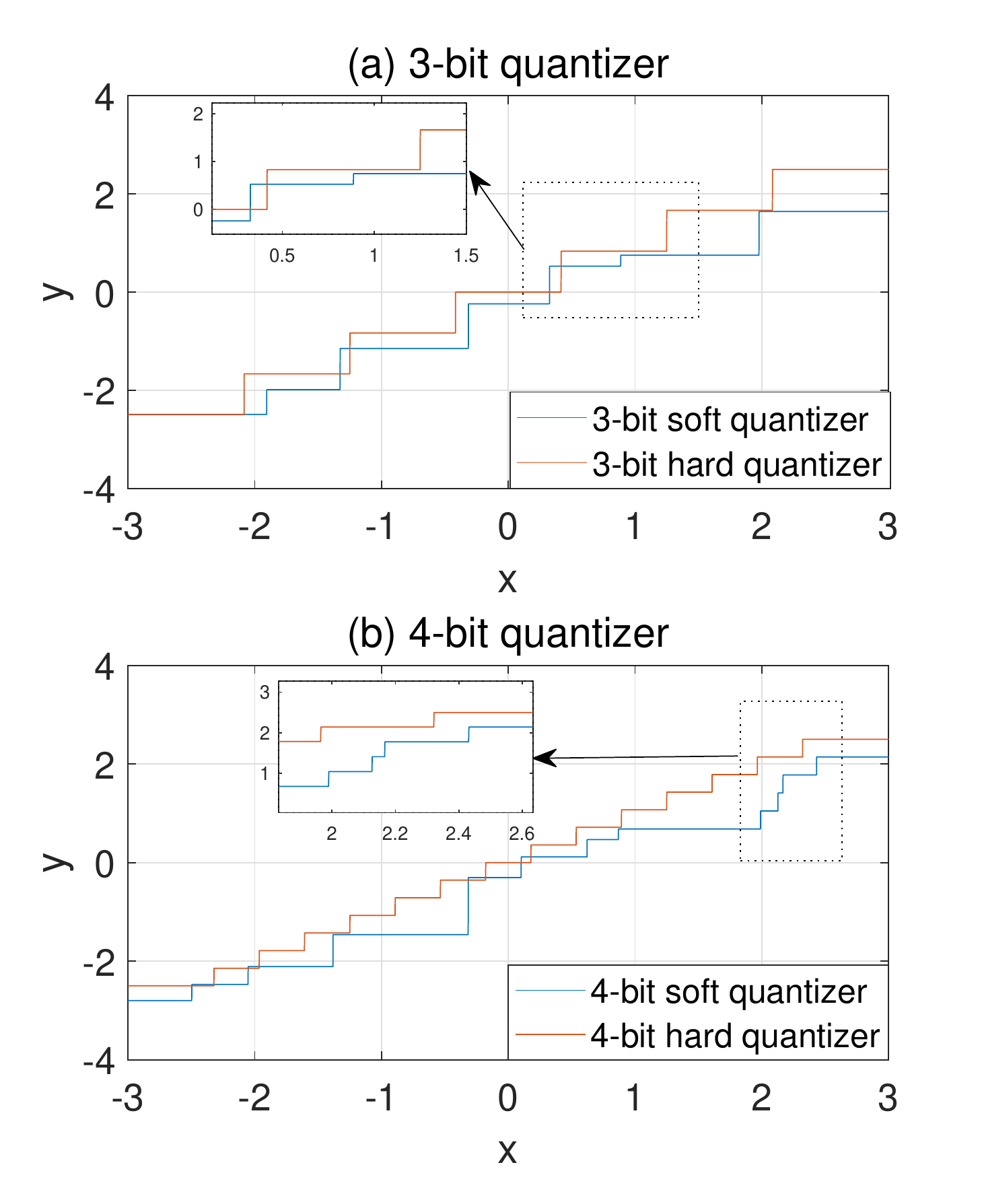}
\caption{Comparison of hard and soft quantizers with different numbers of bits.}
\label{fig9}
\normalsize
\end{figure}

Fig. \ref{fig9} (a) and (b) depicts the staircase functions obtained by the 3-bits/4-bits hard and soft quantizers. From these two subfigures, we can see that the staircase function corresponds to the soft quantizer is nonuniform and its length and height are trained to fit the unquantized parameters ${\bm{\Theta}}_v$.
Besides, we can observe that when the number of quantization bits is $4$, some stairs are automatically merged into a single one and some are broken down into multiple stairs. It seems that the proposed QLcgNetV is trying to distinguish and recognize the importance of each network parameter. Also, compared with the hard quantizer, the number of stairs are reduced from 15 to 12 in this case, which shows the efficiency of the proposed quantizer.
%This phenomenon implies that the number of quantization levels can be automatically adjusted even decided by DL with adequate training data.
%\begin{figure}[h]
%\vspace{2cm}
%\setlength{\belowcaptionskip}{-0.5cm}
%\renewcommand{\captionfont}{\small}
%\centering
%\includegraphics[scale=.50]{bit4quan.eps}
%\caption{Comparison of $4$-bit hard and soft quantizers.}
%\label{fig10}
%\normalsize
%\end{figure}
\section{Conclusion}
In this paper, we proposed a novel model-driven DL network structure, i.e., LcgNet, to address the fundamental massive MIMO detection problem.
LcgNet is essentially designed by unfolding the iterations of the CG detector, where the differences lie in the step-sizes which are discovered to be universal and can be learned through offline training. The dimensions of the step-sizes can be augmented to further improve the detection performance. Moreover, since only a few learnable parameters are required to be optimized, the proposed networks are easy and fast to train.
By inheriting the power of the CG detector and DL, the proposed network showed significant performance gain over the LMMSE detector with much lower computational complexity and this performance gain is achievable under various channel models.

In addition, we presented a novel quantized LcgNetV, i.e., QLcgNetV, where a low-resolution nonuniform quantizer is integrated into LcgNetV to smartly quantize the step-sizes therein. The quantizer was designed by introducing some learnable parameters to a specially designed soft staircase function. Simulation results showed that QLcgNetV can effectively reduce the memory costs with minor detection performance loss.
\section{Extensions}
In this section, we will reveal more potential applications of our proposed LcgNet and present some possible directions for future research.

Note that the CG algorithm is one of the most widely-used optimization algorithms in many practical applications. Besides the massive MIMO detection problem, the proposed LcgNet can also be employed to address many other problems, e.g., low-PAPR precoding design for massive multiuser MIMO systems \cite{7811286} and robust adaptive beamforming for MIMO systems \cite{7558146}, etc.

This work can be viewed as an initial attempt to construct a model-driven DL network by unfolding the CG algorithm, and we validated that the performance of CG  can be further improved by combining state-of-the-art DL methods.  There are many interesting directions to pursue based on this idea. For instance, some variants of the CG algorithm may also be improved by transforming their iterations into data flow graphs and training the resulting unfolded DNNs with a large number of training data, e.g., the biconjugate gradient method (BiCG) \cite{Tapan1987}, the preconditioned CG algorithm for solving singular systems \cite{Erik1988}, projected CG for interior method \cite{Carpenter1993}, some nonlinear CG algorithms such as the Polak-Ribi¡§¡§re-Polyak (PRP) CG algorithm \cite{LZhang2006} and the Dai-Yuan CG algorithms \cite{DaiYuan2000}, etc.

\section{Acknowledgements}
The authors would like to thank Dr. Mingyi Hong from University of Minnesota for his valuable suggestions and criticism which improve the quality of this work.

\bibliographystyle{IEEEtran}
\bibliography{detection}
\end{document}